\documentclass{ws-rv9x6}
\usepackage{ws-rv-thm}   
\usepackage[square]{ws-rv-van}   

\usepackage{feynmp}
\usepackage{shuffle}

\def\beq{\begin{equation}}
\def\eeq{\end{equation}}
\def\bsp#1\esp{\begin{split}#1\end{split}}

\newcommand{\eps}{\epsilon}
\newcommand{\li}[1]{\textrm{Li}_{#1}}
\newcommand{\ord}{\mathcal{O}}
\newcommand{\cS}{\mathcal{S}}
\newcommand{\cI}{\mathcal{I}}

\begin{document}


\chapter*{Mathematical aspects of scattering amplitudes}

\author[C. Duhr]{Claude Duhr}

\address{Center for Cosmology, Particle Physics and Phenomenology,\\
Universit\'e catholique de Louvain,\\
2, Chemin du Cyclotron, 1348 Louvain-La-Neuve, Belgium,\\
Email: claude.duhr@uclouvain.be
}

\begin{abstract}
In these lectures we discuss some of the mathematical structures that appear when computing multi-loop Feynman integrals. We focus on a specific class of special functions, the so-called multiple polylogarithms, and discuss introduce their Hopf algebra structure. We show how these mathematical concepts are useful in physics by illustrating on several examples how these algebraic structures are useful to perform analytic computations of loop integrals, in particular to derive functional equations among polylogarithms.
\end{abstract}
\body

\section{Introduction}
\label{duhr_sec:intro}
The Standard Model of particles physics has been extremely successful in describing experimental data at an unprecedented level of precision. 
When computing predictions for physical observables in the Standard Model, or in any other quantum field theory, a key role is played by scattering amplitudes, which, loosely speaking, encode the differential probability for a certain scattering process to happen. In perturbation theory scattering amplitudes can be expanded into a sum over Feynman diagrams, which at each order involve Feynman graphs with a fixed number of loops. The virtual particles inside the loops are unobservable, and so we need to integrate over their momenta. The computation of perturbative scattering amplitudes beyond tree level therefore necessarily involves the computation of loop integrals. 

Despite the importance of loop Feynman diagrams for precision predictions in quantum field theory, the computation of the corresponding loop integrals is often still a bottleneck. The reasons for this are manifold. For example, loop integrals are Lorentz-invariant functions of the momenta of the external particles in the process, and so multi-leg amplitudes give rise to functions depending on a large number of variables. Moreover, these functions will in general not be elementary functions (say, rational or algebraic), but the functions have a complicated branch cut structure dictated by unitarity and describing intermediate virtual particles going on shell. 

The main purpose of these lectures is to study loop integrals from a purely mathematical and algebraic point of view.
%
%
To be more concrete, we consider in these lectures scalar Feynman integrals of the form
\beq\label{eq:feyn_int}
I=\int\left(\prod_{j=1}^Le^{\gamma_E\eps}\frac{d^Dk_j}{i\pi^{D/2}}\right)\,\frac{\mathcal{N}(\{p_i,k_j\})}{(q_1^2-m_1^2+i0)^{\nu_1}\ldots(q_N^2-m_N^2+i0)^{\nu_N}}\,,
\eeq
where $\nu_i\in\mathbb{Z}$ are integers and $m_i \ge0$, $1\le i\le N$ denote the masses of the propagators. We denote the loop momenta by $k_i$, $1\le i\le L$ and the external momenta by $p_i$, $1\le i\le E$. Note that momentum must be conserved, and in the following we always assume all external momenta in-going, i.e., $\sum_{i=1}^Ep_i=0$. The momenta flowing through the propagators can then be expressed in terms of the loop and external momenta,
\beq
q_i = \sum_{j=1}^L\alpha_{ij}\,k_j + \sum_{j=1}^E\beta_{ij}\,p_j\,,\qquad \alpha_{ij},\beta_{ij}\in \{-1,0,1\}\,.
\eeq
In the following, we will not write the dependence of the propagators on the $+i0$ prescription explicitly any more.
We assume the numerator $\mathcal{N}(\{p_i,k_j\})$ is a polynomial in the scalar products between loop and/or external momenta. We stress that by including numerator factors we also include tensor integrals into the discussion (where all the Lorentz indices have been contracted with suitable external momenta).

We work in dimensional regularisation in $D=D_0-2\eps$ dimensions, $D_0$ a positive integer. We will only consider the case $D_0=4$, although most of what we are going to say also applies to Feynman integrals in other dimensions. It can be shown that $I$ is a \emph{meromorphic function} of $\eps$, i.e., $I$ has at most poles in the complex $\eps$-plane (but no branch cuts!). We will only be interested the Laurent expansion of $I$ close to $\eps=0$, and our main objects of interest will be the coefficients in the Laurent series,
\beq\label{eq:Laurent_exp}
I = \sum_{k\ge k_0}I_k\,\eps^k = I_{k_0}\,\eps^{k_0}+I_{k_0+1}\,\eps^{k_0+1}+I_{k_0+2}\,\eps^{k_0+2}+\ldots\,,\qquad k_0\in\mathbb{Z}\,.
\eeq
If $k_0<0$, $I$ is divergent in $D=D_0$ dimensions.
Note that we include a factor $e^{\gamma_E\eps}/(i\pi^{D/2})$ per loop, where $\gamma_E=-\Gamma'(1)=0.577216\ldots$ is the Euler-Mascheroni constant. The reason for including this normalisation factor will become clear in the next section. 

Feynman integrals, like the one in \eref{eq:feyn_int}, are the main topic of these lectures. More precisely, we will be concerned with the mathematical structure and the properties of the numbers and functions appearing in the analytic expressions for the coefficients of the Laurent expansion~\eqref{eq:Laurent_exp} in dimensional regularisation. The first trivial observation is that Feynman integrals can only depend on Lorentz-invariant quantities, like the scalar products $p_i\cdot p_j$. However, as already pointed out at the beginning of this section the coefficients $I_k$ in \eref{eq:Laurent_exp} may not be simple elementary functions, but rather complicated special functions. These special functions and their properties are the focus of these lectures. 
The main questions we will ask are:
\begin{itemlist}
\item Can any kind of number / function appear in analytic expressions for Feynman integrals?
\item What are the algebraic properties of these functions (functional equations, branch cuts, basis,\ldots)?
\item Can we make general statements about the algebraic and analytic properties of the Laurent coefficients?
\item Can we turn these purely mathematical properties of the functions into concrete tools for Feynman integral computations?
\end{itemlist}

These lectures are organised as follows:
In \sref{duhr_sec:periods} we give a broad classification of the kind of special numbers and functions that can appear in loop computations. In \sref{duhr_sec:MPLs} we introduce the main actors of these lectures, the multiple polylogarithms, and we discuss some of their basic properties. In \sref{duhr_sec:MZVs} and \sref{duhr_sec:Hopf} represent the core of the lectures, and we discuss algebraic and number-theoretical properties underlying these special functions. In \sref{duhr_sec:amps} we give a flavour of how these concepts can be used in loop computations. We include an appendix where we summarise some of the algebraic concepts used throughout the lectures.


\section{Transcendentality and periods}
\label{duhr_sec:periods}
In this section we investigate which classes of numbers and functions can appear in the Laurent coefficients $I_k$. As a warm up, let us look at the following two simple one-loop examples:
\begin{eqnarray}\label{eq:bubble_def}
B(p^2) &=& e^{\gamma_E\eps}\int\frac{d^Dk}{i\pi^{D/2}}\frac{1}{k^2\,(k+p)^2}\,,\\
\label{eq:tri_def}
T(p_1^2,p_2^2,p_3^2) &=& e^{\gamma_E\eps}\int\frac{d^Dk}{i\pi^{D/2}}\frac{1}{k^2\,(k+p_1)^2\,(k+p_1+p_2)^2}\,.
\end{eqnarray}
The integrals are easy to compute, and we get
\begin{eqnarray}\label{eq:bubble_res}
B(p^2) &=&\frac{1}{\eps}+2-\log(-p^2)\\
&&+
\epsilon  \left[\frac{1}{2}\log ^2(-p^2)-2 \log (-p^2)-\frac{1}{2}\,\zeta_2+4\right]
+\ord(\eps^2)\,,\nonumber\\
\label{eq:tri_res}
T(p_1^2,p_2^2,p_3^2) &=&\frac{2}{\sqrt{\lambda}}\,\left[
\li{2}(z) - \li{2}(\bar{z})-\log(z\bar{z})\,\log\frac{1-z}{1-\bar{z}}\right]+\ord(\eps)\,,
\end{eqnarray}
where the variables $z$ and $\bar{z}$ are defined by
\beq
\frac{p_1^2}{p_3^2} = z\bar{z} {\rm~~and~~} \frac{p_2^2}{p_3^2} = (1-z)(1-\bar{z})\,,
\eeq
and $\lambda\equiv \lambda(p_1^2,p_2^2,p_3^2)$ denotes the K\"allen function,
\beq
\lambda(a,b,c)=a^2+b^2+c^2-2ab-2ac-2bc\,.
\eeq
Let us look more closely at Eqs.~\eqref{eq:bubble_res} and~\eqref{eq:tri_res}:  as anticipated, we see that rational functions are insufficient to write down the answer. First, we see the appearance of \emph{zeta values}, i.e., the Riemann $\zeta$ function at integer values,
\beq\label{eq:zeta_n_def}
\zeta_n = \sum_{k=1}^\infty\frac{1}{k^n}\,,\qquad n > 1\,.
\eeq
Note that for $n=1$ the series diverges, and for even $n$, $\zeta_n$ is proportional to $\pi^{2n}$,
\begin{equation}\label{eq:zeta_2n}
\zeta_{2n} = \frac{(-1)^{n+1}\,B_{2n}\,(2\pi)^{2n}}{2\,(2n)!}\,,
\end{equation}
where $B_{2n}$ denote the Bernoulli numbers,
\beq
B_2 =\frac{1}{6}\,,\qquad B_4= -\frac{1}{30}\,,\qquad B_6=\frac{1}{45}\,,\qquad\ldots
\eeq
Next, we see that the answer contains (powers of) \emph{logarithms}
\beq\label{eq:log_def}
\log z = \int_1^z\frac{dt}{t}\,.
\eeq
Moreover, we need generalisations of the logarithm, like the \emph{dilogarithm} $\li{2}(z)$, or more generally, the \emph{classical polylogarithms}, defined recursively by
\beq\label{eq:Li_n_def}
\li{n}(z) = \int_0^z\frac{dt}{t}\li{n-1}(t) = \sum_{k=1}^\infty\frac{z^k}{k^n}\,,
\eeq
where the starting point of the recursion is the ordinary logarithm, $\li{1}(z) = -\log(1-z)$. Note that the series converges for $|z|< 1$. Comparing \eref{eq:zeta_n_def} and \eref{eq:Li_n_def}, we see that
\beq\label{eq:Li_zeta}
\li{n}(1)=\zeta_n\,, \qquad n>1\,.
\eeq
Moreover, we see from \eref{eq:tri_res} that the arguments of the (poly)logarithm are not simple (ratios of) scalar products, but they can be rather complicated functions of the latter.

We see from the previous examples that analytic results for Feynman integrals can easily get pretty involved, already for small numbers of loops and external legs. It is therefore no surprise that for more complicated examples even more complicated functions may arise. In particular, the functions defined in \eref{eq:log_def} and \eref{eq:Li_n_def} are functions of a single argument. In more complicated cases also multi-variable generalisations of the logarithm appear. These functions will be studied in detail in \sref{duhr_sec:MPLs}.  A natural question to ask is: \emph{does the complexity of the functions involved grow indefinitely, i.e., can every function a priori appear in some Laurent coefficient of some Feynman integral?} To be more concrete, we may ask the following questions:
\begin{arabiclist}
\item Can arbitrarily complicated functions appear, e.g., trigonometric functions, exponentials, etc?
\item Can the arguments of these functions be arbitrarily complicated, e.g., could $\log(\log p^2)$ appear?
\item The definition of the integrals in \eref{eq:bubble_def} and~\eqref{eq:tri_def} involves the numbers $e$, $\gamma_E$ and $\log\pi$ (via the Taylor expansion of $\pi^{-D/2}$), but they do not appear in the results for the integrals. Is this an accident?
\end{arabiclist}
In the rest of this section, we will give a complete answer to these questions (and the answers will be negative in all cases).

We have already observed that the results for Feynman integrals contain numbers that are not rational (cf. $\zeta_2=\pi^2/6$ in the previous example). We first need a way to distinguish rational numbers from `non-rational ones':
\begin{definition}
\emph{A complex number is called algebraic (over $\mathbb{Q}$) if it is the root of some polynomial with rational coefficients.}
\end{definition}
A complex number that is not algebraic is called \emph{transcendental (over $\mathbb{Q}$)}.
The set of all algebraic numbers is denoted by $\overline{\mathbb{Q}}$. Remarkably, the set of all algebraic numbers forms a \emph{field}, i.e., the inverse of an algebraic number is algebraic, as well as the sum and the product of two algebraic numbers\footnote{See Appendix~\ref{duhr_app:algebra} for a review of various algebraic structures.}.
We can extend this notion from algebraic and transcendental \emph{numbers} to \emph{functions}: A function is algebraic if it is a root of a polynomial with coefficients that are rational functions in the variables. 

\begin{example}
\begin{romanlist}
\item If $q$ is a rational number, then $q$ is also algebraic, because it is the root of the polynomial $z-q$. In other words, we have $\mathbb{Q}\subset\overline{\mathbb{Q}}$.
\item Every $n$-th root of $q\in\mathbb{Q}$ is algebraic, because $\sqrt[n]{q}$ is a root of $z^n-q$. 
\item In particular, all roots of unity are algebraic, including the imaginary unit $i$. In other words, $\overline{\mathbb{Q}}$ contains also complex numbers.
\item The inverse of $\sqrt{2}$ is algebraic, because $1/\sqrt{2} = \sqrt{2}/{2}$ is a root of $4z^2-2$.
\item $\sqrt{x^2+y^2}$ is an algebraic function because $P(x,y,\sqrt{x^2+y^2})=0$, with $P(x,y,z) = x^2+y^2-z^2$.
\end{romanlist}
\end{example}

We have seen examples of algebraic number, but can be also give examples of transcendental numbers? It is easy to see that not every complex number can be algebraic. Indeed, the set of rational numbers is countable, and so there is a countable number of polynomials with rational coefficients. Since every polynomial has a finite number of roots, the set $\overline{\mathbb{Q}}$ is countable, while the set of all complex numbers is not. In practise it is very difficult to prove that a complex number is transcendental. One of the main results about transcendental numbers is the theorem of Hermite-Lindemann:
\begin{theorem}[Hermite-Lindemann]
Let $z$ be a non-zero complex number. Then either $z$ or $e^z$ is transcendental.
\end{theorem}
The theorem of Hermite-Lindemann allows one to prove that many numbers appearing in loop computations are in fact transcendental.
\begin{example}
\begin{romanlist}
\item $e$ is transcendental, because $e=e^1$, and $1$ is algebraic.
\item $\pi$ is transcendental, because $-1=e^{i\pi}$ and $i$ are algebraic.
\item $\pi^n$, and thus $\zeta_{2n}$, are transcendental for all $n$. Indeed, if $\pi^n$ was algebraic, then there would be a polynomial $P(z)$ with rational coefficients with $P(\pi^n)=0$. But then $\pi$ would be a root of the polynomial $Q(z) \equiv P(z^n)$, which is excluded because $\pi$ is transcendental.
\item $\log q$ is transcendental for all $q\in\overline{\mathbb{Q}}$, because $q = e^{\log q}$ is algebraic.
\end{romanlist}
\end{example}
Looking back at our examples~\eqref{eq:bubble_res} and~\eqref{eq:tri_res}, we see that the Laurent coefficients indeed contain transcendental numbers. Note that the theorem of Hermite-Lindemann does not allow us to determine whether the dilogarithm, and in general polylogarithms, are transcendental or not.  They are, nevertheless, commonly assumed to be transcendental (see \sref{duhr_sec:MPLs}). If we specialise to $\zeta$ values, then we have shown above that all even zetas are transcendental (because they are proportional to $\pi^{2n}$). For odd zetas, only very few transcendentality results are known. In particular, the only odd zeta value that is \emph{proven} to be irrational is $\zeta_3$~\cite{Apery}. We can therefore at best emit the following
\begin{conjecture}
All classical polylogarithms as well as all zeta values are transcendental.
\end{conjecture}

The division into algebraic and transcendental numbers are still too crude to give concrete answers to the questions we asked ourselves at the beginning of this section. It is possible to define a class of number that lies `in between' the algebraic and transcendental numbers. These are the so-called \emph{periods}~\cite{periods}:
\begin{definition} \emph{A complex number is a period if both its real and imaginary parts can be written as integrals of an algebraic function with algebraic coefficients over a domain defined by polynomial inequalities with algebraic coefficients.}
\end{definition}
We will see in the example below that every algebraic number is a period, but not every period is algebraic. Moreover, not every transcendental number is a period. Indeed, there is a countable number of periods (because they are defined using algebraic numbers, and there is only a countable number of those), but there is an uncountable number of transcendental numbers. In other words, there are `more' transcendental numbers than there are periods. Moreover, it can be shown that periods form a \emph{ring}, i.e., sums and products of periods are still periods. Inverses of periods are in general not periods. If we denote the ring of periods by $\mathbb{P}$, then we have the inclusion
\beq
\mathbb{Q}\subset\overline{\mathbb{Q}}\subset \mathbb{P}\subset \mathbb{C}\,.
\eeq
\begin{example}
\begin{romanlist}
\item Every algebraic number $q$  is a period, because $q = \int_0^qdx$.
\item The logarithm of an algebraic number $q$ is a period, because $\log q = \int_1^q\frac{dt}{t}$.
\item $\pi=\int_{x^2+y^2\le1}dx\,dy$ is a period.
\item The dilogarithm (and similarly all polylogarithms and all zeta values) are periods for algebraic arguments, because
\beq
\li{2}(z) = \int_{0\le t_2\le t_1\le z}\frac{dt_1\,dt_2}{t_1\,(1-t_2)}\,.
\eeq
\end{romanlist}
\end{example}
In fact, it turns out that most of the numbers `we know' are periods, and it is rather difficult to prove that a number is not a period! Numbers that are conjectured not to be periods are $e$, $\gamma_E$, $1/\pi$, $\log\pi$,\ldots.

We can now state the main result of this section, which will give the answers to all the questions at the beginning of this section:
\begin{theorem}[Bogner, Weinzierl~\cite{Bogner:2007mn}] In the case where all scalar product $p_i\cdot p_j$ are negative or zero, all internal masses positive, and all ratios of invariants algebraic, the coefficients of the Laurent expansion of a Feynman integral are periods.
\end{theorem}
The proof of the theorem is presented in ref.~\cite{Bogner:2007mn}. The idea of the proof is, loosely speaking, that every Feynman integral admits a Feynman parameter representation,
\begin{eqnarray}\label{eq:FeynPar}
I &=& e^{L\gamma_E\eps}\,(-1)^\nu\,\Gamma\left(\nu-L{D\over 2}\right)\\
&\times&\int_0^1\,\prod_{j=1}^N\left(dx_j'\,\frac{x_j^{\nu_j-1}}{ \Gamma(\nu_j)}\right)\,\delta\left(1-\sum_{j\in S}x_j\right)\,\frac{\mathcal{U}^{\nu-(L+1)D/2}}{(-\mathcal{F})^{\nu-LD/2}}\,,\nonumber
\end{eqnarray}
with $\nu=\sum_{i=1}^N\nu_i$ and $S$ is any non-empty subset of $\{1,\ldots,n\}$, and $\mathcal{U}$ and $\mathcal{F}$ are homogenous polynomials in the Feynman parameters that are completely determined by the topology of the Feynman graph. The main observation is that after expansion in $\eps$ (by means of sector decomposition in the case of divergent integrals) \eref{eq:FeynPar} indeed defines order-by-order an integral of a rational function over some domain defined by rational inequalities, and thus a period. There is one caveat, however: \eref{eq:FeynPar} still explicitly depends on $\gamma_E$, which is expected not to be a period. This factor is exactly cancelled by a similar factor coming from the $\eps$ expansion of the $\Gamma$ function appearing in \eref{eq:FeynPar}. Indeed, using the recursion for the $\Gamma$ function, $\Gamma(1+z)=z\,\Gamma(z)$, as well as the formula
\beq
\Gamma(1+L\eps) =\exp\left(-L\gamma_E\eps+\sum_{k=2}^{\infty}\eps^k\,{(-L)^k\over k}\,\zeta_k\right)\,,
\eeq
it is easy to see that the factor $\exp(L\gamma_E\eps)$ cancels.

It is easy to check that the theorem is true for the examples in \eref{eq:bubble_res} and~\eqref{eq:tri_res}. Moreover, the theorem allows us to answer the three questions we asked at the beginning of the section:
\begin{arabiclist}
\item Trigonometric and exponential functions cannot appear in results for Feynman integrals, because $e$ is (expected) not (to be) a period. Note that inverse trigonometric functions are allowed!
\item The arguments of the polylogarithms should not be arbitrarily complicated: for example, $\log(\log p^2)$ would not be a period for algebraic values of $p^2$.
\item It is not a coincidence that the dependence on $\gamma_E$ and $\log\pi$ cancelled. In fact, this normalisation was introduced precisely to make the theorem true. Note that this normalisation factor is related the one absorbed into the renormalised coupling constant in the $\overline{\textrm{MS}}$-scheme.
\end{arabiclist}


\section{Multiple polylogarithms}
\label{duhr_sec:MPLs}
In the previous section we have seen that (the Laurent coefficients of) Feynman integrals evaluate to a restricted set of numbers and functions called periods, and we have given concrete examples of periods that appear in Feynman integral computations: zeta values and polylogarithms. For multi-loop multi-leg integrals depending on many scales it is known that more complicated generalisations of the logarithm function may appear. In this section we define and study one of these generalisation (bearing in mind that this is not yet the end of the story!), the so-called multiple polylogarithms.

\subsection{Definitions}
Similar to the classical polylogarithms defined in \eref{eq:Li_n_def}, multiple polylogarithms (MPLs) can be defined recursively, for $n\geq 0$, via the iterated integral~\cite{Goncharov:1998,Goncharov:2001}
 \beq\label{eq:MPL_def}
 G(a_1,\ldots,a_n;z)=\,\int_0^z\,{d t\over t-a_1}\,G(a_2,\ldots,a_n;t)\,,\\
\eeq
with $G(z) = G(;z)=1$ and with $a_i\in \mathbb{C}$ are chosen constants and $z$ is a complex variable. In the following, we will also consider $G(a_1,\ldots,a_n;z)$ to be functions of $a_1, \ldots, a_n$.  In the special case where all the $a_i$'s are zero, we define, using the obvious vector notation $\vec a_n=(\underbrace{a,\dots,a}_{n})$, $a\in \mathbb{C}$,
\beq\label{eq:G_0_def}
G(\vec 0_n;z) = {1\over n!}\,\log^n z\,,
\eeq
consistent with the case $n=0$ above. The vector $\vec a=(a_1, \ldots, a_n)$ is called the {\em vector of singularities}  
of the MPL and the number of elements $n$, counted with multiplicities, in that vector is called the {\em weight} of the MPL. Note that the definition of MPLs makes it clear that they are periods (for algebraic values of the arguments). In general, it is not known if they are transcendental, but in the following we will always assume that they are.

\Eref{eq:G_0_def} shows that MPLs contain the ordinary logarithm and the classical polylogarithms as special cases. In particular, we have
\beq
G(\vec a_n;z) = \frac{1}{n!}\log^n\left(1-\frac{z}{a}\right) {\rm~~and~~}
G(\vec 0_{n-1},1;z) = -\li{n}(z)\,.
\eeq
In the case where the $a_i$'s are constant, the above definition was already present in the works of Poincar\'e, Kummer and Lappo-Danilevsky~\cite{Lap35} as ``hyperlogarithms", as well as implicitly in the 1960's in  Chen's work on iterated integrals~\cite{Chen}. Note that the notation for MPLs in the mathematics literature differs slightly from the one used in the physics literature,
\beq\label{eq:math_convention}
I(a_0;a_1,\ldots,a_n;a_{n+1}) = \int_{a_0}^{a_{n+1}}{d t\over t-a_{n}}\,I(a_0;a_1,\ldots,a_{n-1};t)\,,
\eeq
and $I(a_0;;a_1)=1$.
The functions defined by \eref{eq:MPL_def} and \eref{eq:math_convention} are related by (note the reversal of the arguments)
\beq
G(a_n,\ldots,a_1;a_{n+1}) = I(0;a_1,\ldots,a_n;a_{n+1})\,.
\eeq
The iterated integrals defined in \eref{eq:math_convention} are slightly more general than the ones usually defined by physicists, as they allow us to freely choose the base point of the integration. It is nevertheless easy to convert every integral with a generic base point $a_0$ into a combination of iterated integrals with base point $0$.

\begin{example}
It is easy to see that at weight one we have
\beq
I(a_0;a_1;a_2) = I(0;a_1;a_2) - I(0;a_1;a_0) = G(a_1;a_2) - G(a_1;a_0)  \,.
\eeq
Starting from weight two the relation is more complicated because the integrations are nested, {e.g.},
\begin{eqnarray}
I(a_0;a_1,a_2;a_3)& =& \int_{a_0}^{a_3}{d t\over t-a_2}\,I(a_0;a_1;t) \nonumber\\
&=& \int_{a_0}^{a_3}{d t\over t-a_2}\,[I(0;a_1;t) - I(0;a_1;a_0)]\nonumber\\
&=&I(0;a_1,a_2;a_3) - I(0;a_1,a_2;a_0)\\
&& - I(0;a_1;a_0)[I(0;a_2;a_3) - I(0;a_2;a_0)]\nonumber\\
&=&G(a_2,a_1;a_3) - G(a_2,a_1;a_0)\nonumber\\
&& - G(a_1;a_0)[G(a_2;a_3) - G(a_2;a_0)]\nonumber\,.
\end{eqnarray}
\end{example}

In \eref{eq:Li_n_def} we gave two definitions for the classical polylogarithms: a recursive integral definition and a series definition, and the MPLs so far only generalise the integral definition. There is also a way to generalise the series definition~\cite{Goncharov:1998}:
\begin{eqnarray}\label{eq:Li_m_def}
\textrm{Li}_{m_1,\ldots,m_k}(z_1,\ldots,z_k) &=& \sum_{0<n_1<n_2<\dots <n_k} \frac{z_1^{n_1} z_2^{n_2} \cdots z_k^{n_k} }{n_1^{m_1} n_2^{m_2} \cdots n_k^{m_k} }\\
& =&
 \sum_{n_k=1}^\infty{z_k^{n_k}\over n_k^{m_k}}\,\sum_{n_{k-1}=1}^{n_k-1}\ldots\sum_{n_1=1}^{n_{2}-1}{z_1^{n_1}\over n_1^{m_1}}\,,\nonumber
\end{eqnarray}
where this definition makes sense whenever the sums converge (e.g., for $|z_i|<1$). The number $k$ of indices is called the \emph{depth} of the MPL.
Note that we follow Goncharov's original summation convention~\cite{Goncharov:1998}. Other authors define $\textrm{Li}_{m_1,\ldots,m_k}(z_1,\ldots,z_k)$  using the reverse summation convention instead, i.e.~$n_1>\dots>n_k$.
The $G$ and Li functions define (essentially) the same class of functions and are related by
\beq\bsp\label{eq:Gm_def}
\textrm{Li}&_{m_1,\ldots,m_k}(z_1,\ldots,z_k) \\
&\,= (-1)^k\,G\Big(\underbrace{0,\ldots,0}_{m_k-1},{1\over z_k}, \ldots, \underbrace{0,\ldots,0}_{m_1-1},{1\over z_1\ldots z_k};1\Big)\,.
\esp\eeq

\subsection{Basic properties of MPLs}
In this section we discuss some basic properties of MPLs. A large collections of properties (including elementary proofs) can be found in ref.~\cite{Goncharov:2001}. 

It can easily be checked from the integral representation~\eqref{eq:MPL_def} of MPLs that $G(a_1,\ldots,a_n;z)$ is divergent whenever $z=a_1$. Similarly, $G(a_1,\ldots,a_n;z)$ is analytic at $z=0$ whenever $a_n\neq 0$. Note that this is consistent with the series representation, \eref{eq:Li_m_def}. 

If we consider the $a_i$'s constant, then, due to the singularities at $z=a_i$ in the integral representation, multiple polylogarithm will have a very complicated branch cut structure. In particular, $G(a_1,\ldots,a_n;z)$ has branch cuts in the complex $z$ at most extending from $z=a_i$ to $z=\infty$. Note that if the $a_i$'s are allowed to vary, the branch cut structure becomes much more complicated. 
\begin{example}
\begin{romanlist}
\item $G(\vec a_n;z) = \frac{1}{n!}\log^n\left(1-\frac{z}{a}\right)$ has a single branch cut in the complex $z$ plane, extending from $z=a$ to $z=\infty$.
\item $G(0,1;z)=-\li{2}(z)$ has a branch cut extending in the complex $z$ plane from $z=1$ to $z=\infty$. The branch cut starting at $z=0$ is absent in this case.
\end{romanlist}
\end{example}

If the (rightmost) index $a_n$ of $\vec a$  is non-zero, then the function $G(\vec a;x)$ is {\em invariant under a rescaling} of all its arguments, i.e., for any $k\in \mathbb{C}^*$ we have
\beq\label{eq:Gscaling}
G(k\,\vec a;k\, z) = G(\vec a;z)\,,\qquad a_n\neq 0\,.
\eeq
Furthermore, multiple polylogarithms satisfy the {\em H\"older convolution}~\cite{borwein-2001-353}, i.e., whenever $a_1\neq 1$ and $a_n\neq0$, we have, $\forall p\in\mathbb{C}^*$,
\beq\bsp\label{eq:Hoelder}
G&(a_1,\ldots, a_n;1) \\
&\,= \sum_{k=0}^n(-1)^k\,G\left(1-a_k,\ldots, 1-a_1;1-{1\over p}\right)\,G\left(a_{k+1},\ldots, a_n;{1\over p}\right)\,.
\esp\eeq
In the limiting case $p\to \infty$, this identity becomes,
\beq\label{eq:Hoelder_inf}
G(a_1,\ldots, a_n;1) = (-1)^n\,G\left(1-a_n,\ldots, 1-a_1;1\right)\,.
\eeq
The previous examples make it clear that there are many relations among MPLs. Such relations among special functions of the same type are called \emph{functional equations}. While many functional equations among classical polylogarithms can be found in the literature, almost no results are known for functional equations among MPLs. In \sref{duhr_sec:Hopf} we present a way to derive, or rather to circumvent, functional equations among MPLs.

\subsection{The shuffle algebra}
In this section we derive one of the main properties of MPLs (actually, of iterated integrals in general), namely we will see that the product of two MPLs  defined with the same integration limits can be written as a linear combination of MPLs. 

Let us illustrate this in detail on some example, and let us consider the product of two MPLs of weight one. Using the integral representation, we can write
\beq
G(a;z)\,G(b;z) = \int_0^z\frac{dt_1}{t_1-a}\,\int_0^z\frac{dt_2}{t_2-b}
=\iint_{\square}\frac{dt_1\,dt_2}{(t_1-a)\,(t_2-b)}\,,
\eeq
where in the last step we used Fubini's theorem to combine the two integrals into a double integral over the square with corners $(0,0)$, $(0,z)$, $(z,0)$ and $(z,z)$. We can split the square along the diagonal into a sum of two triangles, and we obtain,
\beq\bsp
G&(a;z)\,G(b;z) \\
&\,=\iint_{0\le t_2\le t_1\le z}\frac{dt_1\,dt_2}{(t_1-a)\,(t_2-b)} 
+\iint_{0\le t_1\le t_2\le z}\frac{dt_1\,dt_2}{(t_1-a)\,(t_2-b)}\\
&\,=\int_0^z\frac{dt_1}{t_1-a}\int_0^{t_1}\frac{dt_2}{t_2-b}
+\int_0^z\frac{dt_2}{t_2-b}\int_0^{t_2}\frac{dt_1}{t_1-a}\\
&\,=G(a,b;z)+G(b,a;z)\,.
\esp\eeq
We see that the product of two MPLs of weight one becomes a linear combination of MPLs of weight two. We can repeat exactly the same argument for MPLs of higher weights: we interpret the product of the two integrals as an integral over a hypercube, and recursively split along the diagonals to decompose the integral over the hypercube into iterated integrals. Note that, just like in the example above, it is important that the limits of integration are the same. In the end, we see that a product of MPLs with weights $n_1$ and $n_2$ can always be written as a sum of MPLs with weight $n_1+n_2$,
\beq\bsp\label{eq:G_shuffle}
  G(a_1,\ldots,a_{n_1};z) &\, G(a_{n_1+1},\ldots,a_{n_1+n_2};z) \\
  &\,=\sum_{\sigma\in\Sigma(n_1, n_2)}\,G(a_{\sigma(1)},\ldots,a_{\sigma(n_1+n_2)};z),\\
      \esp\eeq
where $\Sigma(n_1,n_2)$ denotes the set of all \emph{shuffles} of $n_1+n_2$ elements, i.e., the subset of the symmetric group $S_{n_1+n_2}$ defined by (cf. ref.~\cite{Chen})
\beq\bsp\label{eq:Sigma_def}
\Sigma(n_1,n_2) = \{\sigma\in S_{n_1+n_2} &|\, \sigma^{-1}(1)<\ldots<\sigma^{-1}({n_1})\\
 &{\rm and~~} \sigma^{-1}(n_1+1)<\ldots<\sigma^{-1}(n_1+{n_2})\}\,,
\esp\eeq
i.e., the subset of $S_{n_1+n_2}$ that preserves the ordering inside the vectors $(a_1,\ldots,a_{n_1})$ and $(a_{n_1+1},\ldots,a_{n_1+n_2})$. This property turns the set of all MPLs into a \emph{shuffle algebra}, i.e., a vector space equipped with the shuffle multiplication. Note that the shuffle product preserves the weight of the MPLs. We say in this case that the algebra is \emph{graded}.

\begin{example}
\begin{eqnarray}
G(a,b;z)\,G(c;z) &=& G(a,b,c;z) + G(a,c,b;z)+G(c,a,b;z)\,,\\
G(a,b,c;z)\,G(d;z) &=& G(a,b,c,d;z) + G(a,b,d,c;z)\\
&&+G(a,d,b,c;z)+G(d,a,b,c;z)\,,\nonumber\\
G(a,b;z)\,G(c,d;z) &= &G(a,b,c,d;z) + G(a,c,b,d;z)\\
&&+G(c,a,b,d;z)+G(a,c,d,b;z)\nonumber\\
&&+G(c,a,d,b;z)+G(c,d,a,b;z)\nonumber\,.
\end{eqnarray}
\end{example}

\begin{example}\label{ex:shuffle_reg}
In the previous section we have seen that $G(a_1,\ldots,a_n;z)$ is analytic at $z=0$, provided that $a_n\neq 0$. If $a_n=0$, it is always possible to use the shuffle algebra to write $G(a_1,\ldots,a_n;z)$ in terms of functions whose rightmost index of the vector of singularities is non-zero (apart from objects of the form $G(\vec 0_n;x)$), e.g., if $a\neq 0$,
\beq\bsp
G(a,0,0;z) &\,= G(0,0;z)\,G(a;z) - G(0,0,a;z) - G(0,a,0;z)\\
&\, = G(0,0;z)\,G(a;z) - G(0,0,a;z) \\
&\,\phantom{=}\,\,- \left[G(0,a;z)\,G(0;z) - 2G(0,0,a;z)\right]\\
&\,= G(0,0;z)\,G(a;z) + G(0,0,a;z) -G(0,a;z)\,G(0;z)\,.
\esp\eeq
\end{example}

\subsection{The stuffle algebra}
In the previous section we showed that MPLs form a shuffle algebra. This algebra structure is a consequence of the iterated integral definition~\eqref{eq:MPL_def}. In this section we show that there is another algebra structure defined on MPLs, this time induced by the sum representation of MPLs as nested sum,~\eqref{eq:Li_m_def}.

Let us consider the product of two MPLs of depth one. We get
\beq\bsp
\li{1}(z_1)\,\li{1}(z_2) &\,= \sum_{n_1=1}^\infty\frac{z_1^{n_1}}{n_1}\,\sum_{n_2=1}^\infty\frac{z_2^{n_2}}{n_2}\\
&\,=\sum_{n_1,n_2\ge 1}\frac{z_1^{n_1}z_2^{n_2}}{n_1\,n_2}\\
&\,=\left(\sum_{n_2>n_1\ge 1}+\sum_{n_1=n_2\ge 1}+\sum_{n_1>n_2\ge 1}\right)\frac{z_1^{n_1}z_2^{n_2}}{n_1\,n_2}\\
&\,=\li{1,1}(z_1,z_2)+\li{1,1}(z_2,z_1)+\li{2}(z_1z_2)\,.
\esp\eeq
Products of MPLs of higher depths can be handled in a similar way. The algebra generated in this way is called a \emph{stuffle algebra} or \emph{quasi-shuffle algebra}. Just like the shuffle product, the stuffle product preserves the weight. However, it does not preserve the depth, but rather the depth of the product is \emph{bounded} by the sum of the depths. We talk in this case of an algebra \emph{filtered} by the depth. We emphasise that the stuffle algebra structure is completely independent of the shuffle algebra.

\begin{example}
\begin{eqnarray}
\li{m_1,m_2}(z_1,z_2) \,\li{m_3}(z_3) & = &  \li{m_1, m_2, m_3}(z_1, z_2, z_3) \\
+  \li{m_1, m_3, m_2}(z_1, z_3, z_2)
 &+&\li{m_3, m_1, m_2}(z_3, z_1, z_2) \nonumber\\
 +\li{m_1, m_2+m_3}(z_1, z_2z_3)
 &+& \li{m_1+m_3, m_2}(z_1z_3, z_2)\nonumber
 \,,\\
 &&\nonumber\\
 \li{m_1,m_2,m_3}(z_1,z_2,z_3) \,\li{m_4}(z_4) & = &
 \text{Li}_{{m_1},{m_2},{m_3},{m_4}}\left(z_1,z_2,z_3,z_4\right)\\
 \,\,+\text{Li}_{{m_1},{m_2},{m_4},{m_3}}\left(z_1,z_2,z_4,z_3\right)&+&\text{Li}_{{m_1},{m_4},{m_2},{m_3}}\left(z_1,z_4,z_2,z_3\right)\nonumber\\
 \,\,+\text{Li}_{{m_4},{m_1},{m_2},{m_3}}\left(z_4,z_1,z_2,z_3\right)&+&
 \text{Li}_{{m_1},{m_2},{m_3}+{m_4}}\left(z_1,z_2,z_3 z_4\right)\nonumber\\
 +\text{Li}_{{m_1},{m_2}+{m_4},{m_3}}\left(z_1,z_2 z_4,z_3\right)&
 +&\text{Li}_{{m_1}+{m_4},{m_2},{m_3}}\left(z_1 z_4,z_2,z_3\right)\nonumber\,,\\
 &&\nonumber\\
 \li{m_1,m_2}(z_1,z_2)\, \li{m_3,m_4}(z_3,z_4) &= &\text{Li}_{m_1,m_2,m_3,m_4}\left(z_1,z_2,z_3,z_4\right)\\
 +\text{Li}_{m_1,m_3,m_2,m_4}\left(z_1,z_3,z_2,z_4\right)&+&\text{Li}_{m_1,m_3,m_4,m_2}\left(z_1,z_3,z_4,z_2\right)\nonumber\\
 +\text{Li}_{m_3,m_1,m_2,m_4}\left(z_3,z_1,z_2,z_4\right)&+&\text{Li}_{m_3,m_1,m_4,m_2}\left(z_3,z_1,z_4,z_2\right)\nonumber\\
 +\text{Li}_{m_3,m_4,m_1,m_2}\left(z_3,z_4,z_1,z_2\right)&+&
 \text{Li}_{m_1+m_3,m_2+m_4}\left(z_1 z_3,z_2 z_4\right)\nonumber\\
 +\text{Li}_{m_1,m_3,m_2+m_4}\left(z_1,z_3,z_2 z_4\right)&+&\text{Li}_{m_1,m_2+m_3,m_4}\left(z_1,z_2 z_3,z_4\right)\nonumber\\
 +\text{Li}_{m_3,m_1,m_2+m_4}\left(z_3,z_1,z_2 z_4\right)&+&\text{Li}_{m_3,m_1+m_4,m_2}\left(z_3,z_1 z_4,z_2\right)\nonumber\\
 +\text{Li}_{m_1+m_3,m_2,m_4}\left(z_1 z_3,z_2,z_4\right)&+&\text{Li}_{m_1+m_3,m_4,m_2}\left(z_1 z_3,z_4,z_2\right)\nonumber\,.
 \end{eqnarray}
 \end{example}

 \subsection{Special instances of MPLs}
 \label{duhr_subsec:GHPLs}
 Multiple polylogarithms are a very general class of functions that contain many other functions as special cases. In particular, there are several classes of special functions introduced by physicists in the context of specific Feynman integral computations that can be expressed through MPLs. In this section we give a brief review of these functions, which commonly appear in loop computations. 

\begin{arabiclist}
{\bf \item Harmonic polylogarithms (HPLs)~\cite{Remiddi:1999ew}:} HPLs correspond to the special case where $a_i\in\{-1,0,1\}$. For historical reasons, harmonic polylogarithms only agree with MPLs up to a sign, and are denoted by $H$ rather than $G$. The exact relation between HPLs and MPLs is
\beq
H(\vec a;z) = (-1)^p\,G(\vec a;z)\,,
\eeq
where $p$ is the number of elements in the vector $\vec a$ equal to $(+1)$. Because of the importance of HPLs for phenomenology, they have been implemented into various computer codes that allow one to evaluate HPLs numerically in a fast and reliable way~\cite{Gehrmann:2001pz,Maitre:2005uu,Maitre:2007kp,Vollinga:2004sn,Buehler:2011ev}.
{\bf \item Two-dimensional harmonic polylogarithms (2dHPLs)~\cite{Gehrmann:2000zt}:} 2dHPLs correspond to the special case where $a_i\in\{0,1,-y,-1-y\}$, for $y\in\mathbb{C}$. They appear in the computation of four-point functions with three on-shell and one off shell leg~\cite{Gehrmann:2000zt,Gehrmann:2001ck,DiVita:2014pza}, and can be evaluated numerically using the techniques of ref.~\cite{Vollinga:2004sn,Gehrmann:2001jv}.
{\bf \item Generalized harmonic polylogarithms (GHPLs)~\cite{Aglietti:2004tq}:} GHPLs are defined as iterated integrals involving square roots of quadratic polynomials as integration kernels, e.g.,
\beq\label{eq:GHPL_def}
G(-r,\vec a;z) = \int_0^z\frac{dt}{\sqrt{t(4+t)}}\,G(\vec a;t)\,,
\eeq
whenever the integral converges. These integrals appear in loop amplitudes that present a two-particle threshold at $s=4m^2$ ($z=-s/m^2$). In ref.~\cite{Bonciani:2010ms} it was shown that GHPLs can always be expressed in terms of MPLs via the change of variable
\beq\label{eq:GHPL_change}
z = \frac{(1-\xi)^2}{\xi}\,,\qquad \xi = \frac{\sqrt{4+z}-\sqrt{z}}{\sqrt{4+z}+\sqrt{z}}\,.
\eeq
Letting $t=(1-\eta)^2/\eta$ in \eref{eq:GHPL_def}, we find
\beq
G(-r,\vec a;z) = -\int_1^\xi\frac{d\eta}{\eta}\,G\left(\vec a;\frac{(1-\eta)^2}{\eta}\right)\,.
\eeq
If we assume recursively that the $G$-function in the right-hand side can be expressed through MPLs of the form $G(\ldots;\eta)$, then it is easy to see that the remaining integral will lead to MPLs.
\begin{example} Let us consider $G(-r,-1;z)$. Performing the change of variables~\eqref{eq:GHPL_change}, we get,
\beq
G(-r,-1;z) = -\int_1^\xi\frac{d\eta}{\eta}\,G\left(-1;\frac{(1-\eta)^2}{\eta}\right)\,.
\eeq
We have
\beq\bsp
G\left(-1;\frac{(1-\eta)^2}{\eta}\right) &\,= \log\left(1+\frac{(1-\eta)^2}{\eta}\right)\\
&\, = \log(1-\eta+\eta^2)-\log\eta\\
&\,=\log(1-c\,\eta)+\log(1-\bar{c}\,\eta)-\log\eta\\
&\,=G(\bar{c};\eta) + G(c;\eta)-G(0;\eta)\,,
\esp\eeq
\end{example}
where $c=\exp(i\pi/3)$ and $\bar{c}=\exp(-i\pi/3)$ are two primitive sixth roots of unity.
So we get
\beq\bsp
G&(-r,-1;z) \\
&\,= -G(0,\bar{c};\xi)-G(0,c;\xi)+G(0,0;\xi)+G(0,\bar{c};1)+G(0,c;1)\,.
\esp\eeq
{\bf \item Cyclotomic harmonic polylogarithms (CHPLs)~\cite{Ablinger:2011te}:} CHPLs are a generalisation of HPLs defined by the iterated integrals
\beq
C^{a\vec l}_{b\vec m}(z) = \int_0^zdt\,f_b^a(t)\,C^{\vec l}_{\vec m}(t)\,,
\eeq
with 
\beq
f_0^0(z) = \frac{1}{z} {\rm~~and~~} f_m^l(z) = \frac{z^l}{\Phi_m(z)}\,, \quad 0\le l\le \varphi(m)\,,
\eeq
where $\phi(m)$ is Euler's totient function and $\Phi_m(z)$ denotes the $m$-th cyclotomic polynomial,
\beq
\Phi_1(z) = z-1\,,\quad \Phi_2(z) = z+1\,,\quad \Phi_3(z) = z^2+z+1\,,\quad\ldots
\eeq
By definition, the roots of cyclotomic polynomials are roots of unity. Thus, if we factor the cyclotomic polynomials and us partial fractioning, we can express all CHPLs in terms of MPLs where the $a_i$'s are roots of unity.
\begin{example} Let us consider $C_6^0(z)$. we have
\beq\bsp
C_6^0(z) &\,= \int_0^z\frac{dt}{\Phi_6(t)} = \int_0^z\frac{dt}{1-t+t^2}
=\int_0^z\frac{dt}{(t-c)(t-\bar{c})}\\
&\,=\frac{1}{c-\bar{c}}\left[G(c;z)-G(\bar{c};z)\right]
=-\frac{i}{\sqrt{3}}\left[G(c;z)-G(\bar{c};z)\right]\,,
\esp\eeq
with $c=\exp(i\pi/3)$ and $\bar{c}=\exp(-i\pi/3)$.
\end{example}

\end{arabiclist}


\section{Multiple zeta values}
\label{duhr_sec:MZVs}
 
\subsection{Definition of MZVs}

In \eref{eq:Li_zeta} we saw that there is a connection between classical polylogarithms and the values of the Riemann zeta function at positive integers. It is natural to ask how to generalise these relations to the more general polylogarithmic functions defined in \sref{duhr_sec:MPLs}.

In this section we discuss \emph{multiple zeta values (MZVs)}, a `multi-index' extension of the ordinary zeta values defined in \eref{eq:zeta_n_def}. \emph{Ordinary} zeta values are the values at $1$ of \emph{classical} polylogarithms, and so it is natural to define MZVs as the values at 1 of MPLs,
\begin{definition}
\emph{Let $m_1,\ldots,m_k$ be positive integers. 
\beq\label{eq:MZV_def}
\zeta_{m_1,\ldots,m_k} = \li{m_k,\ldots,m_1}(1,\ldots,1) = \sum_{n_1>\ldots>n_k>0}{1\over {n_1^{m_1}\ldots n_k^{m_k}}}\,.
\eeq}
\end{definition}
Note that if $m_1=1$ in \eref{eq:MZV_def}, then $\zeta_{1,m_2,\ldots,m_k}$ is divergent. In the following we will only consider convergent series. The weight and the depth of an MZV are defined in the same way as for MPLs.
The reason to study these numbers (just like ordinary zeta values, MZV will be numbers, not functions!) is twofold: First, they are ubiquitous in both mathematics and in multi-loop computations, and so they deserve a deeper study. Second, they allow us to introduce some of the concepts and the way of thinking that we will use in subsequent sections, but in a simpler and more controlled framework. As such, this section also serves as a preparation for subsequent sections.

Using \eref{eq:Gm_def}, we see that MZVs also admit a definition in terms of iterated integrals, and the integral is convergent whenever the MZV is. Note that the existence of this integral representation implies that all MZVs are periods (`integrals of rational functions'). In \sref{duhr_sec:periods} we already mentioned that, apart from $\zeta_3$, it is not known if a given odd zeta value is transcendental or not, and it is therefore not surprising that basically nothing is known about the transcendentality of MZVs. In the following we will assume the `usual folklore' that
\begin{conjecture}\label{conj:MZV1}
All MZVs are transcendental.
\end{conjecture}

\subsection{Relations among MZVs}
In \sref{duhr_sec:periods} we have seen that all ordinary even zeta values are proportional to powers of $\pi^2$, or in other words, all even ordinary zeta values are related, $\zeta_{2n}=c\, \zeta_2^n$, for some $c\in\mathbb{Q}$. The main question we will ask ourselves in the rest of this section is whether there are more such relations among MZVs. Actually, we already have at our disposal a machinery to generate infinite numbers of relations! We know that MPLs satisfy shuffle and stuffle relations, and so by listing systematically all the shuffle and stuffle relations among (convergent) MZVs, we can generate lots of relations. 
\begin{example}
\begin{romanlist}
\item Using the stuffle algebra, we can write
\beq\label{eq:stuffle_4}
\zeta_2^2 = \li{2}(1)^2=2\li{1,1}(1,1)+\li{4}(1) = 2\zeta_{2,2}+\zeta_4\,.
\eeq
Similarly, using the shuffle algebra,
\beq\bsp\label{eq:shuffle_4}
\zeta_2^2&\,=G(0,1;1)^2 = 4G(0,0,1,1;1) + 2 G(0,1,0,1;1)\\
&\, = 4\zeta_{3,1}+2\zeta_{2,2}\,.
\esp\eeq
\item Using the stuffle algebra, we can write
\beq\bsp
\zeta_2\,\zeta_3 &\,= \li{2}(1)\,\li{3}(1)=\li{2,3}(1,1)+\li{3,2}(1,1)+\li{5}(1)\\
&\, = \zeta_{2,3}+\zeta_{3,2}+\zeta_5\,.
\esp\eeq
Similarly, using the shuffle algebra,
\beq\bsp
\zeta_2\,\zeta_3&\,=G(0,1;1)\,G(0,0,1;1) \\
&\,= 6G(0,0,0,1,1;1)+3G(0,0,1,0,1;1) + G(0,1,0,0,1;1) \\
&\,= 6\zeta_{4,1}+3\zeta_{2,3}+\zeta_{3,2}\,.
\esp\eeq
\end{romanlist}
\end{example}
We see that, because the shuffle and stuffle products preserve the weight, we can only generate relations among MZVs of the same weight in this way. Note that the first time we can generate shuffle or stuffle identities is at weight four, because at lower weight all products involve divergent MZVs. There are, however, relations among MZVs of weight three
\beq\label{eq:z12}
\zeta_{2,1} = \zeta_3\,.
\eeq
In other words, there must be more exotic relations among MZVs than the shuffle and stuffle relations among convergent MZVs we have considered so far.
This brings up the following interesting questions:
\begin{arabiclist}
\item Are there relations among MZVs of different weight?
\item Can we characterise \emph{all} the relations that exist between MZVs?
\item Can we describe a `basis' for MZVs, at least for fixed weight?
\end{arabiclist}
Amazingly, all of these questions can be answered, at least at the level of conjectures (that have been tested numerically to hundreds of digits for rather high weights). 

In order to formulate these conjectures, and also to prepare the ground for the following sections, let us rephrase these questions in mathematical language. First, we need to be slightly more precise and define what kind of relations we are looking for. In the following, we mean by `relations among MZVs' a relation of the type $P(Z_1,\ldots,Z_n) = 0$, where $Z_i$ are MZVs and $P$ is a polynomial with rational coefficients. The first conjecture states that
\begin{conjecture}\label{conj:MZV2}
There are no relations among MZVs of different weights.
\end{conjecture}
This implies that all the terms in the polynomial $P$ have the same weight.
Note that our definition of `relation' relies crucially on our Conjecture~\ref{conj:MZV1} that all MZVs are transcendental! Indeed, suppose that there is an MZV $Z_0$ of weight $n_0$ that is a rational number. Then for any other MZV $Z_1$ of weight $n_1$ we could write $P(Z_0,Z_1)=0$, where $P(x,y) = Z_0^{-1}\,xy-y=0$ is a polynomial with rational coefficients. In other words, we would have obtained a relation among MZVs of weight $n_0+n_1$ and $n_1$.

Let us now denote the vector space of all convergent MZVs of weight $n>1$ by $\mathcal{Z}_n$. By definition, we put $\mathcal{Z}_0=\mathbb{Q}$ and $\mathcal{Z}_1=\{0\}$ (because there are no convergent MZVs of weight one). Furthermore, we define the vector space of all MZVs to be the direct sum of all the $\mathcal{Z}_n$,
\beq
\mathcal{Z} = \bigoplus_{n=0}^\infty\mathcal{Z}_n = \mathbb{Q} \oplus \mathcal{Z}_2\oplus\mathcal{Z}_3 \oplus \ldots\,.
\eeq
This definition might look innocent at first glance, but it contains a lot of deep mathematical statements! In particular, it embodies already the conjectures~\ref{conj:MZV1} and~\ref{conj:MZV2}. Indeed, the fact that the sum is direct implies that $\mathcal{Z}_m\cap\mathcal{Z}_n=\{0\}$, for $m\neq n$. If $Z_0\in\mathcal{Z}_n$, $n\neq 0$, is rational, then $Z_0\in\mathcal{Z}_0\cap\mathcal{Z}_n=\{0\}$, and so $Z_0=0$. In other words, there is no rational MZV. Similarly, assume that there is a relation between MZVs of different weights, say $m$ and $n$. This means that there are elements $Z_1\in\mathcal{Z}_m$ and $Z_2\in\mathcal{Z}_n$ such that $Z_1+Z_2=0$, and so $Z_1=-Z_2$. But then $Z_1$, $Z_2\in\mathcal{Z}_m\cap\mathcal{Z}_n=\{0\}$, and so $Z_1=Z_2=0$.

Next, note that $\mathcal{Z}$ is actually not only a vector space, but it is an algebra, because the MZVs can be equipped with a product (say, the shuffle product). We have already seen in \sref{duhr_sec:MPLs} that the shuffle product preserves the weight, and we called such an algebra \emph{graded}. We can now formalise this by saying that whenever $Z_1\in\mathcal{Z}_m$ and $Z_2\in \mathcal{Z}_n$, we have $Z_1Z_2\in\mathcal{Z}_{m+n}$.

We can now formulate the previous questions in our new language of vector spaces:
\begin{arabiclist}
\item What are the dimensions of the vector spaces $\mathcal{Z}_n$?
\item Can we write down an explicit basis for each of the $\mathcal{Z}_n$?
\end{arabiclist}
We can answer these questions (at least conjecturally) using the celebrated \emph{double-shuffle conjecture}. Loosely speaking, the conjecture states that, if we formally also include the \emph{divergent} MZVs, then the only relations among the \emph{convergent} MZVs are those that can be obtained via shuffle and stuffle identities.
\begin{example} 
If we formally include all divergent MZVs, we can write the following stuffle relation at weight three:
\beq\bsp\label{eq:z1z2_stuffle}
\zeta_1\,\zeta_2 &\,= \li{1}(1)\,\li{2}(1) = \li{2,1}(1,1)+\li{1,2}(1,1)+\li{3}(1) \\
&\,= \zeta_{1,2}+\zeta_{2,1}+\zeta_3\,.
\esp\eeq
Similarly, we can write the shuffle relation
\beq\bsp\label{eq:z1z2_shuffle}
\zeta_1\,\zeta_2&\,=G(1;1)\,G(0,1;1) = G(1,0,1;1) + 2 G(0,1,1;1)\\
&\, = \zeta_{1,2}+2\zeta_{2,1}\,.
\esp\eeq
Note that these relations are purely formal, because both sides of the equalities are divergent. However, both in~\eref{eq:z1z2_stuffle} and~\eref{eq:z1z2_shuffle} the only divergent quantity in the right-hand side is $\zeta_{1,2}$. If we take the difference of the two equations, then all the divergent quantities cancel, and we are left with a concrete relation among convergent MZVs:
\beq
0 = -\zeta_{2,1} + \zeta_3\,,
\eeq
i.e., we find obtained the `exotic' relation~\eqref{eq:z12}. In other words, we have obtained a relation between convergent MZVs as the difference between (formal) shuffle and stuffle identities. Such a relation is called a \emph{regularised shuffle relation}.
\end{example}
\begin{conjecture}
The only relations among MZVs are shuffle, stuffle and regularised shuffle relations.
\end{conjecture}
\begin{example}
We have already derived shuffle and stuffle relations at weight four in \eref{eq:stuffle_4} and~\eqref{eq:shuffle_4}. We can now add regularised shuffle relations. We start by writing formal stuffle relations
\begin{eqnarray}
\zeta_1\,\zeta_3 &=& \zeta_{1,3}+\zeta_{3,1}+\zeta_4\,,\\
\zeta_1\,\zeta_{2,1} &=& \zeta_{1,2,1}+2\zeta_{2,1,1}+\zeta_{2,2}+\zeta_{3,1}\,,
\end{eqnarray}
and shuffle relations
\begin{eqnarray}
\zeta_1\,\zeta_3 &=& \zeta_{1,3}+2\zeta_{3,1}+\zeta_{2,2}\,,\\
\zeta_1\,\zeta_{2,1} &=& \zeta_{1,2,1}+3\zeta_{2,1,1}\,.
\end{eqnarray}
Taking the difference, we obtain two regularised identities among convergent MZVs of weight four,
\begin{eqnarray}
0 &=& -\zeta_{3,1}+\zeta_4-\zeta_{2,2}\,,\\
0 &=& -\zeta_{2,1,1} + \zeta_{2,2} + \zeta_{3,1}\,.
\end{eqnarray}
Combining these relations with \eref{eq:stuffle_4} and~\eqref{eq:shuffle_4}, we have obtained four relations among MVZs of weight 4. The solution is
\beq
\zeta_4 = \zeta_{2,1,1} = \frac{2}{5}\zeta_2^2\,, \qquad \zeta_{3,1} = \frac{1}{10}\zeta_2^2\,,\qquad \zeta_{2,2}=\frac{3}{10}\zeta_{2}^2\,,
\eeq
i.e., all MZVs of weight four are proportional to $\pi^4$!
\end{example}

The double-shuffle conjecture answers the two-questions we asked earlier, because we can, at least in principle, solve the double-shuffle relations for each weight, and in this way we can construct an explicit basis for each $\mathcal{Z}_n$. This has been done explicitly up to high weights in ref.~\cite{Blumlein:2009cf}. Moreover, there is a conjecture about the dimensions $d_n = \textrm{dim}_{\mathbb{Q}}\mathcal{Z}_n$:
\beq
d_1=0 \,,\qquad d_0=d_2=1,\qquad d_k = d_{k-2}+d_{k-3}\,,\,\,\,\, k>2\,.
\eeq
In \tref{tab:MZV} we show an explicit basis of MZVs up to weight eight. Note that the first time a generic MZV can no longer be written as a polynomial in ordinary zeta values is at weight eight. Moreover, one can give an explicit basis for every weight~\cite{HoffmannBasis,BrownMixedTate}: the MZVs of the form $\zeta_{m_1,\ldots,m_k}$ with $m_i\in\{2,3\}$ are expected to form a basis of all MZVs.

\begin{table}[!t]
\tbl{Basis of MZVs up to weight eight.}
{\begin{tabular}{@{}cccccccc@{}}  \colrule
Weight &  2 & 3 & 4 & 5 & 6 & 7 & 8\\
Dimension& 1 & 1 & 1 & 2 & 2 & 3 & 4 \\
Basis & $\zeta_2$ & $\zeta_3$ & $\zeta_2^2$ & $\zeta_2\zeta_3$, $\zeta_5$ & 
$\zeta_2^3$, $\zeta_3^2$ & $\zeta_2^2\zeta_3$, $\zeta_2\zeta_5$, $\zeta_7$ & 
$\zeta_2^4$, $\zeta_2\zeta_3^2$, $\zeta_3\zeta_5$, $\zeta_{5,3}$\\
\botrule
\end{tabular}}
\label{tab:MZV}
\end{table}

\section{The Hopf algebra of MPLs}
\label{duhr_sec:Hopf}

\subsection{Functional equations among MPLs}
In the previous section we have seen that it is possible, at least at the level of conjectures, to describe all the relations among MZVs, and to give a complete basis of MZVs at each weight. In this section we will generalise this idea to the framework of MPLs. It is clear that finding all the relations among MPLs is a monumental task, which is much more complicated than in the case of MZVs. In particular, it is clear that for MPLs there must be new relations that go beyond 
shuffle and stuffle relations, because we now have to deal with \emph{functions} rather than numbers, and so we will also need to take into account relations among MPLs with different arguments. In the rest of these lectures we refer to such relations as \emph{functional equations}. The main question we will try to answer in this section is thus: Is there a way to describe functional equations among MPLs?

In order to get a feeling for functional equations, let us look at some simple representatives: At weight one, there is precisely one fundamental functional equation, namely
\beq
\log(ab) = \log a + \log b\,.
\eeq
All other functional equations for the logarithm are just a consequence of this relation. At weight two, we have for example the following functional equations for the dilogarithm
\begin{eqnarray}
\li{2}(1-z) &=& -\li{2}(z) -\log(1-z)\,\log z+\zeta_2\,,\\
\li{2}\left(1-\frac{1}{z}\right) &=& -\li{2}(1-z) -\frac{1}{2}\log^2z\,,
\end{eqnarray}
as well as the five-term relation
\beq\bsp
\li{2}\left(\frac{x}{1-y}\right)&\, + \li{2}\left(\frac{y}{1-x}\right) - \li{2}\left(\frac{xy}{(1-x)(1-y)}\right) \\
&\,= \li{2}(x)+\li{2}(y) + \log(1-x)\log(1-y)\,.
\esp\eeq
Note that these identities are only valid for specific values of the variables. Many more identities can be found in the literature (see, e.g., ref.~\cite{Lewin,LewinBook1}). 

Functional equations like the ones above are not only interesting from the purely mathematical point of view, but they also play an important role when computing Feynman integrals. For example, polylogarithms have branch cuts, and functional equations can be used to analytically continue the functions. In the previous sections we argued that also MPLs appear in Feynman integral computations. It is then not surprising that functional equations for MPLs are also needed in physics. Unfortunately, not many examples of functional equations for MPLs are known in the mathematics literature. For this reason, there is a substantial literature in physics where functional equations for special classes of MPLs have been studied (cf. \sref{duhr_subsec:GHPLs}), e.g., ref.~\cite{Remiddi:1999ew,Gehrmann:2000zt,Gehrmann:2001jv,Gehrmann:2001pz,Aglietti:2004tq,Bonciani:2010ms,Ablinger:2011te,Ablinger:2010kw,Ablinger:2013cf,Maitre:2005uu,Maitre:2007kp,Vollinga:2004sn}. All of the methods presented in these references are tailored to specific special classes of functions, and usually require the manipulation of the integral or series representations of the functions (e.g., an identity derived via some change of variables in the integral representation).

The purpose of this section is to present a method that allows one to derive functional equations among MPLs (or at least some classes of functional equations). The main differences to the special cases considered so far in the physics literature are
\begin{arabiclist}
\item the method is completely generic and applies to arbitrary MPLs, and is not tailored to specific special classes of iterated integrals.
\item the method is completely algebraic and combinatorial in nature, and it is completely agnostic of the underlying integral or series representations.
\end{arabiclist}

We will use an algebraic framework similar to the one used for MZVs in the previous section. In particular, let us define $\mathcal{A}_n$ as the vector space spanned by all `polylogarithmic functions' of weight $n$, and we put $\mathcal{A}_0=\mathbb{Q}$. Note that $\mathcal{A}_n$ includes all MZVs of weight $n$, $\mathcal{Z}_n\subset\mathcal{A}_n$, but unlike in the MZV-case, $\mathcal{A}_1\neq\{0\}$, because $\mathcal{A}_1$ contains all ordinary logarithms. We also define $\mathcal{A}$ to be the direct sum of the vector spaces $\mathcal{A}_n$,
\beq
\mathcal{A} = \bigoplus_{n=0}^\infty\mathcal{A}_n\,.
\eeq
Just like in the case of MZVs, this definition only makes sense if we assume the following
\begin{conjecture}
All MPLs are transcendental functions, and there are no relations among MPLs of different weights.
\end{conjecture}
Obviously, $\mathcal{A}$ is an algebra, given by the multiplication of functions (cf. the shuffle and stuffle products), and we know already that this algebra is graded by the weight. Unlike the MZV-case, where all relations are given by shuffle and stuffle identities, there are much more complicated relations among MPLs, and those relations cannot be recovered from shuffle and stuffle relations alone. In particular, all the functional equations that change the arguments of the functions cannot be covered by double-shuffles. We therefore need a much more general and flexible algebraic framework if we want to find all the relations among MPLs of a given weight.

\subsection{Coalgebras and Hopf algebras}
In this section we briefly review the algebraic concepts that we will need to formulate our framework. We will not give a detailed account of all the mathematical definitions, and content ourselves to give the basics that are needed to follow the discussions in the remaining sections. More detailed definitions can be found in Appendix~\ref{duhr_app:algebra}.

We have already seen that $\mathcal{A}$ is an algebra, i.e., a vector space with a multiplication that has a unit element and is associative, $(ab)c=a(bc)$, and distributive, $a(b+c)=ab+ac$ and $(a+b)c=ac+bc$. In particular, there is a map, the multiplication, which assigns to a pair of elements $(a,b)$ their product $ab$. It will be useful to see the multiplication as a map $\mu$ from $\mathcal{A}\otimes\mathcal{A}$ to $\mathcal{A}$, and the pair $(a,b)$ will be denoted by $a\otimes b$. We will not make use of all the properties of the tensor product $\mathcal{A}\otimes\mathcal{A}$. Here it suffices to say that $a\otimes b$ behaves just like a pair of elements, subject to the bilinearity conditions
\begin{eqnarray}
&&(a+b)\otimes c = a\otimes c+b\otimes c\,,\qquad a\otimes(b+c)=(a\otimes b)+(a\otimes c)\,,\\
&&(k\,a)\otimes b=a\otimes (k\,b) = k\,(a\otimes b)\,,
\end{eqnarray}
$\forall a,b,c\in\mathcal{A}$ and $k\in\mathbb{Q}$. Moreover, if $\mathcal{A}$ is an algebra, then $\mathcal{A}\otimes\mathcal{A}$ is an algebra as well, and the multiplication is defined `component-wise',
\beq\label{eq:tensor_mult}
(a\otimes b)(c\otimes d) = (ac)\otimes(bd)\,.
\eeq

We need an additional algebraic structure: a \emph{coalgebra} is a vector space $\mathcal{A}$ equipped with a \emph{comultiplication}, i.e., a linear map $\Delta:\mathcal{A}\to\mathcal{A}\otimes \mathcal{A}$ which assigns to every element $a\in\mathcal{A}$ its coproduct $\Delta(a) \in \mathcal{A}\otimes\mathcal{A}$. Moreover, the comultiplication is required to be \emph{coassociative}, $(\Delta\otimes \textrm{id})\Delta) = (\textrm{id}\otimes \Delta)\Delta$. The meaning of the coassiciativity is the following: the coproduct is a prescription that assigns to every element $a\in\mathcal{A}$ a `pair' of elements (or rather, a linear combination of pairs). We can schematically write
\beq
a\mapsto \Delta(a) = \sum_{i}a^{(1)}_i\otimes a^{(2)}_i\,.
\eeq
If we have a prescription to split an element $a$ into two, then we can iterate this prescription to split $a$ into three. However, we can do this in two different ways,
\begin{eqnarray}
a& \mapsto& \sum_{i}a^{(1)}_i\otimes a^{(2)}_i \mapsto \sum_{i}\Delta(a^{(1)}_i)\otimes a^{(2)}_i = \sum_{ij} a^{(1,1)}_{ij}\otimes a^{(1,2)}_{ij}\otimes a^{(2)}_i\,,\\
a& \mapsto &\sum_{i}a^{(1)}_i\otimes a^{(2)}_i \mapsto \sum_{i}a^{(1)}_i\otimes \Delta(a^{(2)}_i) = \sum_{ij} a^{(1)}_{i}\otimes a^{(2,1)}_{ij}\otimes a^{(2,2)}_{ij}\,.
\end{eqnarray}
Coassociativity states that these two expressions are the same, i.e., the order in which we iterate the coproduct is immaterial. In other words, their is are unique prescriptions to split an object into two, three, four, etc. pieces.

Finally, if $\mathcal{A}$ is equipped with both a multiplication and a comultiplication, we require them to be compatible in the sense that the coproduct of a product is the product of the coproducts,
\beq\label{eq:compatibility}
\Delta(ab) = \Delta(a)\,\Delta(b)\,,
\eeq 
where in the right-hand side the multiplication should be interpreted according to \eref{eq:tensor_mult}. A vector space with compatible multiplications and comultiplications is called a \emph{bialgebra}. If the bialgebra is graded as an algebra, we require the coproduct to respect the weight as well, i.e., the sum of the weights of the two factors in the coproduct of $a$ equals the weight of $a$. 

\begin{example}\label{ex:Hopf}
Consider a set of letters, say $\{a,b,c\}$, and let us consider the vector space $\mathcal{A}$ spanned by all linear combinations of words (with rational coefficients) in these letters. There is a natural multiplication on $\mathcal{A}$, given by concatenation of words, e.g., $(ab)\otimes c \mapsto abc$. 
Note that $\mathcal{A}$ is graded, and the weight is given by the length of the word. 
Next, let us define a linear map $\Delta:\mathcal{A}\to\mathcal{A}\otimes\mathcal{A}$ in the following way:
\begin{romanlist}
\item on letters, $\Delta$ acts like $\Delta(x) = 1\otimes x+x\otimes 1$, $x\in\{a,b,c\}$.
\item we extend the definition to words of length $\ge2$ using the compatibility condition~\eqref{eq:compatibility}.
\end{romanlist}
For example, we have
\begin{eqnarray}
\Delta(1) &=&1\otimes 1\,,\\
\Delta(ab) &=& \Delta(a)\Delta(b) = (1\otimes a+a\otimes 1)(1\otimes b+b\otimes 1)\\
&=& 1\otimes (ab) + (ab)\otimes 1+a\otimes b + b\otimes a\,.\nonumber\\
\Delta(abc) &=& \Delta(ab)\Delta(c)\\
& =& (1\otimes (ab) + (ab)\otimes 1+a\otimes b + b\otimes a)(1\otimes c+c\otimes 1)\nonumber\\
&=& 1\otimes (abc)+(abc)\otimes 1+(ab)\otimes c + b\otimes(ac) + c\otimes (ab)\nonumber\\
&& + (ac)\otimes b + a\otimes (bc) + (bc)\otimes a\nonumber\,.
\end{eqnarray}
Note that the coproduct respects the weight, i.e., the length of a word.
Let us explicitly check coassociativity.
We can now iterate the coproduct of $abc$. If we iterate in the first entry, we get
\beq\bsp\label{eq:DID}
(\Delta&\otimes\textrm{id})\Delta(abc) =\Delta(1)\otimes (abc)+\Delta(abc)\otimes 1+\Delta(ab)\otimes c\\
& + \Delta(b)\otimes(ac) + \Delta(c)\otimes (ab) + \Delta(ac)\otimes b + \Delta(a)\otimes (bc)\\
& + \Delta(bc)\otimes a\\
&=1\otimes1\otimes (abc)+
1\otimes (abc)\otimes1+(abc)\otimes 1\otimes1+(ab)\otimes c\otimes1\\
& + b\otimes(ac)\otimes1 + c\otimes (ab)\otimes1 + (ac)\otimes b\otimes1 + a\otimes (bc)\otimes1 \\
&+ (bc)\otimes a\otimes1+
1\otimes (ab)\otimes c + (ab)\otimes 1\otimes c+a\otimes b\otimes c \\
&+ b\otimes a\otimes c+
1\otimes b\otimes(ac) + b\otimes 1\otimes(ac) +
1\otimes c\otimes(ab) \\
&+ c\otimes 1\otimes(ac) +
 1\otimes (ac)\otimes b + (ac)\otimes 1\otimes b+a\otimes c\otimes b \\
 &+ c\otimes a\otimes b+
 1\otimes a\otimes(bc) + a\otimes 1\otimes(bc) +
  1\otimes (bc)\otimes a \\
  &+ (bc)\otimes 1\otimes a+b\otimes c\otimes a + c\otimes b\otimes a\,.
\esp\eeq
Similarly, if we iterate in the second entry, we get
\beq\bsp\label{eq:IDD}
(\textrm{id}&\otimes\Delta)\Delta(abc) =
1\otimes \Delta(abc)+(abc)\otimes \Delta(1)+(ab)\otimes \Delta(c)\\
& + b\otimes\Delta(ac) + c\otimes\Delta(ab) + (ac)\otimes \Delta(b) + a\otimes \Delta(bc) \\
&+ (bc)\otimes \Delta(a)\\
&=1\otimes1\otimes (abc)+1\otimes(abc)\otimes 1+1\otimes(ab)\otimes c +1\otimes b\otimes(ac)\\
& + 1\otimes c\otimes (ab)+ 1\otimes(ac)\otimes b +1\otimes a\otimes (bc) + 1\otimes(bc)\otimes a\\
&+(abc)\otimes 1\otimes 1+
(ab)\otimes 1\otimes c+(ab)\otimes c\otimes 1+
 b\otimes1\otimes (ac) \\
 &+  b\otimes(ac)\otimes 1+ b\otimes a\otimes c +  b\otimes c\otimes a+
 c\otimes1\otimes (ab)\\
 & +  c\otimes(ab)\otimes 1+ c\otimes a\otimes b +  c\otimes b\otimes a+
(ac)\otimes 1\otimes b\\
&+(ac)\otimes b\otimes 1+
  a\otimes1\otimes (bc) +  a\otimes(bc)\otimes 1+ a\otimes b\otimes c \\
  &+  a\otimes c\otimes b+
(bc)\otimes 1\otimes a+(bc)\otimes a\otimes 1\,.
\esp\eeq
We see that \eref{eq:DID} and \eref{eq:IDD} give the same result, i.e., the coproduct is coassociative, and so $\mathcal{A}$ is a bialgebra.
\end{example}

A \emph{Hopf algebra} is a bialgebra together with an additional structure, called \emph{antipode}, that we do not need in the following. We will therefore skip the definition of the antipode here and identify Hopf algebras and bialgebras. 
We conclude this section by introducing some definitions:
\begin{arabiclist}
\item An element $x$ in a Hopf algebra is called \emph{primitive} if $\Delta(x)=1\otimes x+x\otimes 1$, i.e., $x$ is primitive if it cannot be decomposed in any non-trivial way.
\item The \emph{reduced} coproduct is defined by $\Delta'(x) = \Delta(x) - 1\otimes x - x\otimes 1$.
\item If the Hopf algebra is graded, we introduce maps $\Delta_{i_1,\dots,i_k}$ which assign to an element $x$ the the part of the iterated coproduct where the factors in the coproduct have weights $(i_1,\ldots,i_k)$.
\end{arabiclist}

\begin{example} Using the definitions from Example~\ref{ex:Hopf}, we see that all letters are primitive elements. The reduced coproducts are
\begin{eqnarray}
\Delta'(ab) &=& a\otimes b+b\otimes a\,,\\
\Delta'(abc) &=& (ab)\otimes c + b\otimes(ac) + c\otimes (ab) + (ac)\otimes b + a\otimes (bc)\\
&& + (bc)\otimes a\,.\nonumber
\end{eqnarray}
\end{example}
The different components of the coproduct are
\begin{eqnarray}
\Delta_{1,1}(ab) &=& a\otimes b+b\otimes a\,,\\
\Delta_{2,1}(abc) &=& (ab)\otimes c + (bc)\otimes a + (ac)\otimes b\,,\\
\Delta_{1,2}(abc) &=& a\otimes (bc) + b\otimes (ac) + c\otimes (ab)\,,\\
\Delta_{1,1,1}(abc) &=& a\otimes b\otimes c + b\otimes c\otimes a + c\otimes a\otimes b\\
&&+a\otimes c\otimes b + b\otimes a\otimes c + c\otimes b\otimes a\,.\nonumber
\end{eqnarray}

\subsection{The Hopf algebra of MPLs}
In this section we show that MPLs form a Hopf algebra, and we define the coproduct on MPLs~\cite{Goncharov:2005sla}. The construction and the definition of the coproduct is a bit subtle, and it will be carried out in three stages:
\begin{arabiclist}
{\bf \item The coproduct in the generic case:} We start by defining a coproduct on MPLs of the form $I(a_0;a_1,\ldots,a_n;a_{n+1})$, where the $a_i$ are \emph{generic}, i.e., the $a_i$ do not take special values and $a_i\neq a_j$ if $i\neq j$.
{\bf \item Shuffle regularisation:} In a second step, we extend the definition to the non-generic case, where for example some of the $a_i$ are allowed to be equal. This introduces additional singularities that need to be regularised.
{\bf \item Inclusion of even zeta values:} Finally, we show how to consistently include the even zeta values.
\end{arabiclist}

\paragraph{The coproduct in the generic case.} In this section we define the coproduct on $I(a_0;a_1,\ldots,a_n;a_{n+1})$, where the $a_i$ are generic~\cite{Goncharov:2005sla}. It is more convenient to work with the $I$-notation rather than the $G$-notation, because it makes some of the formulas more transparent. The coproduct on MPLs is defined by~\cite{Goncharov:2005sla}
\beq\bsp\label{eq:coproduct}
\Delta&(I(a_0;a_1,\ldots,a_n;a_{n+1})) \\
&= \sum_{0=i_1<i_2<\ldots<i_{k}<i_{k+1}=n} I(a_0;a_{i_1},\ldots,a_{i_k};a_{n+1})\\
&\qquad\otimes\Bigg[\prod_{p=0}^kI(a_{i_p};a_{i_p+1},\ldots,a_{i_{p+1}-1};a_{i_{p+1}})\Bigg]\,.
\esp\eeq
In ref.~\cite{Goncharov:2005sla} it was shown that \eref{eq:coproduct} defines a genuine coproduct, i.e., it is coassociative and compatible with the multiplication. The different terms in \eref{eq:coproduct} admit a simple combinatorial description:
\begin{romanlist}
\item Draw a semi-circle (including the diameter), and mark $(n+2)$ points on the semi-circle by $a_0$, $a_1$, \ldots, $a_n$, $a_{n+1}$, arranged clockwise and such that $a_0$ and $a_{n+1}$ are the end-points of the diameter.
\item Select some of the marked points (including the possibility to select none!), say $a_{i_1}$,\ldots, $a_{i_k}$, $1\le i_l\le n$ and draw the convex polygon with with vertices vertices $a_0$, $a_{i_1}$,\ldots, $a_{i_k}$, $a_{n+1}$. This polygon defines the first factor of a term in \eref{eq:coproduct}.
\item The unmarked points defines a family of complementary convex polygons, e.g., the first complementary polygon has vertices $a_0$, $a_1$, \ldots, $a_{i_1-1}$, $a_{i_1}$, the second complementary polygon has vertices $a_{i_1}$, $a_{i_1+1}$, \ldots, $a_{i_2-1}$, $a_{i_2}$ and so on. These polygons define the MPLs in the product in the second factor of each term in \eref{eq:coproduct}
\end{romanlist}

\begin{example}\label{ex:Delta1}
Let us consider the coproduct of a generic MPL of weight one, $I(a_0;a_1;a_2)$. There are only two different ways to select points on the half circle:
\begin{center}
\begin{tabular}{ccc}
\begin{fmffile}{W1term1}
\begin{fmfgraph*}(110,70)
\fmfbottom{a0,a2}
\fmfdot{a0}
\fmfdot{a2}
\fmfv{label=$a_0$}{a0}
\fmfv{label=$a_2$}{a2}
\fmf{plain}{a0,a2}
\fmffreeze
\fmf{plain,left,tension=0.2,tag=1}{a0,a2}
\fmfposition
\fmfipath{p[]}
\fmfiset{p1}{vpath1(__a0,__a2)}
\fmfiv{d.sh=circle,d.f=1,d.size=2thick,label=$a_1$}{point length(p1)/2 of p1}
\fmffreeze\fmfposition
\fmfipair{a[]}
\fmfiequ{a1}{point length(p1)/2 of p1}
\fmfi{dashes}{vloc(__a0) -- a1}
\fmfi{dashes}{a1 -- vloc(__a2)}
\end{fmfgraph*}
\end{fmffile}
&\phantom{aaaaaa}
&  \begin{fmffile}{W1term2}
\begin{fmfgraph*}(110,70)
\fmfbottom{a0,a2}
\fmfdot{a0}
\fmfdot{a2}
\fmfv{label=$a_0$}{a0}
\fmfv{label=$a_2$}{a2}
\fmf{plain}{a0,a2}
\fmffreeze
\fmf{plain,left,tension=0.2,tag=1}{a0,a2}
\fmfposition
\fmfipath{p[]}
\fmfiset{p1}{vpath1(__a0,__a2)}
\fmfiv{d.sh=circle,d.f=1,d.size=2thick,label=$a_1$}{point length(p1)/2 of p1}
\fmffreeze\fmfposition
\fmfipair{a[]}
\fmfiequ{a1}{point length(p1)/2 of p1}
\fmfi{plain}{vloc(__a0) -- a1}
\fmfi{plain}{a1 -- vloc(__a2)}
\end{fmfgraph*}
\end{fmffile}
 \\
${\displaystyle1\otimes I(a_0;a_1;a_2)}$ && ${\displaystyle I(a_0;a_1;a_2)\otimes 1}$ \\
\end{tabular}
\end{center}
The solid polygon represents the polygon built on the vertices $a_0$, $a_1$, $a_2$; in that case there is no complementary polygon, and so the second factor in the coproduct is just 1. The dashed polygon denotes a complementary polygon with vertices $a_0$, $a_1$, $a_2$ (it is complementary to the trivial polygon with vertices $a_0$ and $a_2$). We see that MPLs of weight one are primitive elements, i.e., they cannot be decomposed further.
\end{example}

\begin{example}\label{ex:Delta2}
Let us consider the coproduct of a generic MPL of weight two, $I(a_0;a_1,a_2;a_3)$. First, there are the two trivial ways to inscribe a polygon into the semi-circle:
\begin{center}
\begin{tabular}{ccc}
\begin{fmffile}{W2term1}
\begin{fmfgraph*}(110,70)
\fmfbottom{a0,a3}
\fmfdot{a0}
\fmfdot{a3}
\fmfv{label=$a_0$}{a0}
\fmfv{label=$a_3$}{a3}
\fmf{plain}{a0,a3}
\fmffreeze
\fmf{plain,left,tension=0.2,tag=1}{a0,a3}
\fmfposition
\fmfipath{p[]}
\fmfiset{p1}{vpath1(__a0,__a3)}
\fmfiv{d.sh=circle,d.f=1,d.size=2thick,label=$a_1$}{point length(p1)/3 of p1}
\fmfiv{d.sh=circle,d.f=1,d.size=2thick,label=$a_2$}{point 2length(p1)/3 of p1}
\fmffreeze\fmfposition
\fmfipair{a[]}
\fmfiequ{a1}{point length(p1)/3 of p1}
\fmfiequ{a2}{point 2length(p1)/3 of p1}
\fmfi{dashes}{vloc(__a0) -- a1}
\fmfi{dashes}{a1 -- a2}
\fmfi{dashes}{a2 -- vloc(__a3)}
\end{fmfgraph*}
\end{fmffile}
&\phantom{aaaaaa}
&\begin{fmffile}{W2term2}
\begin{fmfgraph*}(110,70)
\fmfbottom{a0,a3}
\fmfdot{a0}
\fmfdot{a3}
\fmfv{label=$a_0$}{a0}
\fmfv{label=$a_3$}{a3}
\fmf{plain}{a0,a3}
\fmffreeze
\fmf{plain,left,tension=0.2,tag=1}{a0,a3}
\fmfposition
\fmfipath{p[]}
\fmfiset{p1}{vpath1(__a0,__a3)}
\fmfiv{d.sh=circle,d.f=1,d.size=2thick,label=$a_1$}{point length(p1)/3 of p1}
\fmfiv{d.sh=circle,d.f=1,d.size=2thick,label=$a_2$}{point 2length(p1)/3 of p1}
\fmffreeze\fmfposition
\fmfipair{a[]}
\fmfiequ{a1}{point length(p1)/3 of p1}
\fmfiequ{a2}{point 2length(p1)/3 of p1}
\fmfi{plain}{vloc(__a0) -- a1}
\fmfi{plain}{a1 -- a2}
\fmfi{plain}{a2 -- vloc(__a3)}
\end{fmfgraph*}
\end{fmffile}
 \\
${\displaystyle1\otimes I(a_0;a_1,a_2;a_3)}$ && ${\displaystyle I(a_0;a_1,a_2;a_3)\otimes 1}$ \\
\end{tabular}
\end{center}
The dashed line represents the complementary polygon. These two terms represent the two trivial terms, and are the analogues of the two terms encountered in Example~\ref{ex:Delta1}. At weight two we have two additional terms,
\begin{center}
\begin{tabular}{ccc}
\begin{fmffile}{W2term3}
\begin{fmfgraph*}(110,70)
\fmfbottom{a0,a3}
\fmfdot{a0}
\fmfdot{a3}
\fmfv{label=$a_0$}{a0}
\fmfv{label=$a_3$}{a3}
\fmf{plain}{a0,a3}
\fmffreeze
\fmf{plain,left,tension=0.2,tag=1}{a0,a3}
\fmfposition
\fmfipath{p[]}
\fmfiset{p1}{vpath1(__a0,__a3)}
\fmfiv{d.sh=circle,d.f=1,d.size=2thick,label=$a_1$}{point length(p1)/3 of p1}
\fmfiv{d.sh=circle,d.f=1,d.size=2thick,label=$a_2$}{point 2length(p1)/3 of p1}
\fmffreeze\fmfposition
\fmfipair{a[]}
\fmfiequ{a1}{point length(p1)/3 of p1}
\fmfiequ{a2}{point 2length(p1)/3 of p1}
\fmfi{plain}{vloc(__a0) -- a1}
\fmfi{plain}{a1 -- vloc(__a3)}
\fmfi{dashes}{a1 -- a2}
\fmfi{dashes}{a2 -- vloc(__a3)}
\end{fmfgraph*}
\end{fmffile}
&\phantom{aaaaaa}
&\begin{fmffile}{W2term4}
\begin{fmfgraph*}(110,70)
\fmfbottom{a0,a3}
\fmfdot{a0}
\fmfdot{a3}
\fmfv{label=$a_0$}{a0}
\fmfv{label=$a_3$}{a3}
\fmf{plain}{a0,a3}
\fmffreeze
\fmf{plain,left,tension=0.2,tag=1}{a0,a3}
\fmfposition
\fmfipath{p[]}
\fmfiset{p1}{vpath1(__a0,__a3)}
\fmfiv{d.sh=circle,d.f=1,d.size=2thick,label=$a_1$}{point length(p1)/3 of p1}
\fmfiv{d.sh=circle,d.f=1,d.size=2thick,label=$a_2$}{point 2length(p1)/3 of p1}
\fmffreeze\fmfposition
\fmfipair{a[]}
\fmfiequ{a1}{point length(p1)/3 of p1}
\fmfiequ{a2}{point 2length(p1)/3 of p1}
\fmfi{plain}{vloc(__a0) -- a2}
\fmfi{plain}{a2 -- vloc(__a3)}
\fmfi{dashes}{vloc(__a0) -- a1}
\fmfi{dashes}{a1 -- a2}
\end{fmfgraph*}
\end{fmffile}
 \\
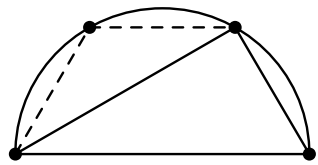
${\displaystyle I(a_0;a_1;a_3)\otimes I(a_1;a_2;a_3)}$ && ${\displaystyle I(a_0;a_2;a_3)\otimes I(a_0;a_1;a_2)}$ \\
\end{tabular}
\end{center}
\end{example}

\begin{example}\label{ex:Delta3}
At weight three, we have the usual two trivial terms:
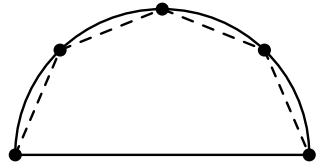
\begin{center}
\begin{tabular}{ccc}
\begin{fmffile}{W3term1}
\begin{fmfgraph*}(110,70)
\fmfbottom{a0,a4}
\fmfdot{a0}
\fmfdot{a4}
\fmfv{label=$a_0$}{a0}
\fmfv{label=$a_4$}{a4}
\fmf{plain}{a0,a4}
\fmffreeze
\fmf{plain,left,tension=0.2,tag=1}{a0,a4}
\fmfposition
\fmfipath{p[]}
\fmfiset{p1}{vpath1(__a0,__a4)}
\fmfiv{d.sh=circle,d.f=1,d.size=2thick,label=$a_1$}{point length(p1)/4 of p1}
\fmfiv{d.sh=circle,d.f=1,d.size=2thick,label=$a_2$}{point 2length(p1)/4 of p1}
\fmfiv{d.sh=circle,d.f=1,d.size=2thick,label=$a_3$}{point 3length(p1)/4 of p1}
\fmffreeze\fmfposition
\fmfipair{a[]}
\fmfiequ{a1}{point length(p1)/4 of p1}
\fmfiequ{a2}{point 2length(p1)/4 of p1}
\fmfiequ{a3}{point 3length(p1)/4 of p1}
\fmfi{dashes}{vloc(__a0) -- a1}
\fmfi{dashes}{a1 -- a2}
\fmfi{dashes}{a2 -- a3}
\fmfi{dashes}{a3 -- vloc(__a4)}
\end{fmfgraph*}
\end{fmffile}
&\phantom{aaaaaa}
&\begin{fmffile}{W3term2}
\begin{fmfgraph*}(110,70)
\fmfbottom{a0,a4}
\fmfdot{a0}
\fmfdot{a4}
\fmfv{label=$a_0$}{a0}
\fmfv{label=$a_4$}{a4}
\fmf{plain}{a0,a4}
\fmffreeze
\fmf{plain,left,tension=0.2,tag=1}{a0,a4}
\fmfposition
\fmfipath{p[]}
\fmfiset{p1}{vpath1(__a0,__a4)}
\fmfiv{d.sh=circle,d.f=1,d.size=2thick,label=$a_1$}{point length(p1)/4 of p1}
\fmfiv{d.sh=circle,d.f=1,d.size=2thick,label=$a_2$}{point 2length(p1)/4 of p1}
\fmfiv{d.sh=circle,d.f=1,d.size=2thick,label=$a_3$}{point 3length(p1)/4 of p1}
\fmffreeze\fmfposition
\fmfipair{a[]}
\fmfiequ{a1}{point length(p1)/4 of p1}
\fmfiequ{a2}{point 2length(p1)/4 of p1}
\fmfiequ{a3}{point 3length(p1)/4 of p1}
\fmfi{plain}{vloc(__a0) -- a1}
\fmfi{plain}{a1 -- a2}
\fmfi{plain}{a2 -- a3}
\fmfi{plain}{a3 -- vloc(__a4)}
\end{fmfgraph*}
\end{fmffile}
 \\
${\displaystyle1\otimes I(a_0;a_1,a_2,a_3;a_4)}$ && ${\displaystyle I(a_0;a_1,a_2,a_3;a_4)\otimes 1}$ \\
\end{tabular}
\end{center}
Next, we have five non-trivial terms with one complementary polygon each:
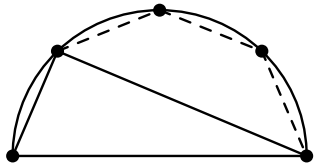
\begin{center}
\begin{tabular}{ccc}
\begin{fmffile}{W3term3}
\begin{fmfgraph*}(110,70)
\fmfbottom{a0,a4}
\fmfdot{a0}
\fmfdot{a4}
\fmfv{label=$a_0$}{a0}
\fmfv{label=$a_4$}{a4}
\fmf{plain}{a0,a4}
\fmffreeze
\fmf{plain,left,tension=0.2,tag=1}{a0,a4}
\fmfposition
\fmfipath{p[]}
\fmfiset{p1}{vpath1(__a0,__a4)}
\fmfiv{d.sh=circle,d.f=1,d.size=2thick,label=$a_1$}{point length(p1)/4 of p1}
\fmfiv{d.sh=circle,d.f=1,d.size=2thick,label=$a_2$}{point 2length(p1)/4 of p1}
\fmfiv{d.sh=circle,d.f=1,d.size=2thick,label=$a_3$}{point 3length(p1)/4 of p1}
\fmffreeze\fmfposition
\fmfipair{a[]}
\fmfiequ{a1}{point length(p1)/4 of p1}
\fmfiequ{a2}{point 2length(p1)/4 of p1}
\fmfiequ{a3}{point 3length(p1)/4 of p1}
\fmfi{plain}{vloc(__a0) -- a1}
\fmfi{plain}{a1 -- vloc(__a4)}
\fmfi{dashes}{a1 -- a2}
\fmfi{dashes}{a2 -- a3}
\fmfi{dashes}{a3 -- vloc(__a4)}
\end{fmfgraph*}
\end{fmffile}
&\phantom{aaaaaa}
&\begin{fmffile}{W3term4}
\begin{fmfgraph*}(110,70)
\fmfbottom{a0,a4}
\fmfdot{a0}
\fmfdot{a4}
\fmfv{label=$a_0$}{a0}
\fmfv{label=$a_4$}{a4}
\fmf{plain}{a0,a4}
\fmffreeze
\fmf{plain,left,tension=0.2,tag=1}{a0,a4}
\fmfposition
\fmfipath{p[]}
\fmfiset{p1}{vpath1(__a0,__a4)}
\fmfiv{d.sh=circle,d.f=1,d.size=2thick,label=$a_1$}{point length(p1)/4 of p1}
\fmfiv{d.sh=circle,d.f=1,d.size=2thick,label=$a_2$}{point 2length(p1)/4 of p1}
\fmfiv{d.sh=circle,d.f=1,d.size=2thick,label=$a_3$}{point 3length(p1)/4 of p1}
\fmffreeze\fmfposition
\fmfipair{a[]}
\fmfiequ{a1}{point length(p1)/4 of p1}
\fmfiequ{a2}{point 2length(p1)/4 of p1}
\fmfiequ{a3}{point 3length(p1)/4 of p1}
\fmfi{dashes}{vloc(__a0) -- a1}
\fmfi{dashes}{a1 -- a2}
\fmfi{dashes}{a2 -- a3}
\fmfi{plain}{vloc(__a0) -- a3}
\fmfi{plain}{a3 -- vloc(__a4)}
\end{fmfgraph*}
\end{fmffile}
 \\
${\displaystyle I(a_0;a_1;a_4)\otimes I(a_1;a_2,a_3;a_4)}$ && ${\displaystyle I(a_0;a_3;a_4)\otimes  I(a_0;a_1,a_2;a_3)}$ \\
\end{tabular}
\end{center}
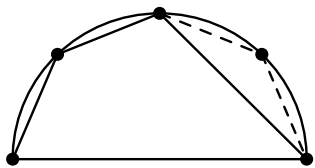
\begin{center}
\begin{tabular}{ccc}
\begin{fmffile}{W3term5}
\begin{fmfgraph*}(110,70)
\fmfbottom{a0,a4}
\fmfdot{a0}
\fmfdot{a4}
\fmfv{label=$a_0$}{a0}
\fmfv{label=$a_4$}{a4}
\fmf{plain}{a0,a4}
\fmffreeze
\fmf{plain,left,tension=0.2,tag=1}{a0,a4}
\fmfposition
\fmfipath{p[]}
\fmfiset{p1}{vpath1(__a0,__a4)}
\fmfiv{d.sh=circle,d.f=1,d.size=2thick,label=$a_1$}{point length(p1)/4 of p1}
\fmfiv{d.sh=circle,d.f=1,d.size=2thick,label=$a_2$}{point 2length(p1)/4 of p1}
\fmfiv{d.sh=circle,d.f=1,d.size=2thick,label=$a_3$}{point 3length(p1)/4 of p1}
\fmffreeze\fmfposition
\fmfipair{a[]}
\fmfiequ{a1}{point length(p1)/4 of p1}
\fmfiequ{a2}{point 2length(p1)/4 of p1}
\fmfiequ{a3}{point 3length(p1)/4 of p1}
\fmfi{plain}{vloc(__a0) -- a1}
\fmfi{plain}{a1 -- a2}
\fmfi{plain}{a2 -- vloc(__a4)}
\fmfi{dashes}{a2 -- a3}
\fmfi{dashes}{a3 -- vloc(__a4)}
\end{fmfgraph*}
\end{fmffile}
&\phantom{aaaaaa}
&\begin{fmffile}{W3term6}
\begin{fmfgraph*}(110,70)
\fmfbottom{a0,a4}
\fmfdot{a0}
\fmfdot{a4}
\fmfv{label=$a_0$}{a0}
\fmfv{label=$a_4$}{a4}
\fmf{plain}{a0,a4}
\fmffreeze
\fmf{plain,left,tension=0.2,tag=1}{a0,a4}
\fmfposition
\fmfipath{p[]}
\fmfiset{p1}{vpath1(__a0,__a4)}
\fmfiv{d.sh=circle,d.f=1,d.size=2thick,label=$a_1$}{point length(p1)/4 of p1}
\fmfiv{d.sh=circle,d.f=1,d.size=2thick,label=$a_2$}{point 2length(p1)/4 of p1}
\fmfiv{d.sh=circle,d.f=1,d.size=2thick,label=$a_3$}{point 3length(p1)/4 of p1}
\fmffreeze\fmfposition
\fmfipair{a[]}
\fmfiequ{a1}{point length(p1)/4 of p1}
\fmfiequ{a2}{point 2length(p1)/4 of p1}
\fmfiequ{a3}{point 3length(p1)/4 of p1}
\fmfi{dashes}{vloc(__a0) -- a1}
\fmfi{dashes}{a1 -- a2}
\fmfi{plain}{vloc(__a0) -- a2}
\fmfi{plain}{a2 -- a3}
\fmfi{plain}{a3 -- vloc(__a4)}
\end{fmfgraph*}
\end{fmffile}
 \\
${\displaystyle I(a_0;a_1,a_2;a_4)\otimes I(a_2;a_3;a_4)}$ && ${\displaystyle I(a_0;a_2,a_3;a_4)\otimes  I(a_0;a_1;a_2)}$ \\
\end{tabular}
\end{center}
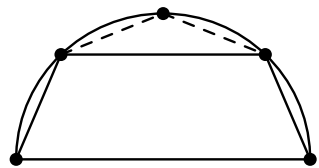
\begin{center}
\begin{tabular}{c}
\begin{fmffile}{W3term7}
\begin{fmfgraph*}(110,70)
\fmfbottom{a0,a4}
\fmfdot{a0}
\fmfdot{a4}
\fmfv{label=$a_0$}{a0}
\fmfv{label=$a_4$}{a4}
\fmf{plain}{a0,a4}
\fmffreeze
\fmf{plain,left,tension=0.2,tag=1}{a0,a4}
\fmfposition
\fmfipath{p[]}
\fmfiset{p1}{vpath1(__a0,__a4)}
\fmfiv{d.sh=circle,d.f=1,d.size=2thick,label=$a_1$}{point length(p1)/4 of p1}
\fmfiv{d.sh=circle,d.f=1,d.size=2thick,label=$a_2$}{point 2length(p1)/4 of p1}
\fmfiv{d.sh=circle,d.f=1,d.size=2thick,label=$a_3$}{point 3length(p1)/4 of p1}
\fmffreeze\fmfposition
\fmfipair{a[]}
\fmfiequ{a1}{point length(p1)/4 of p1}
\fmfiequ{a2}{point 2length(p1)/4 of p1}
\fmfiequ{a3}{point 3length(p1)/4 of p1}
\fmfi{plain}{vloc(__a0) -- a1}
\fmfi{plain}{a1 -- a3}
\fmfi{plain}{a3 -- vloc(__a4)}
\fmfi{dashes}{a1 -- a2}
\fmfi{dashes}{a2 -- a3}
\end{fmfgraph*}
\end{fmffile}\\
${\displaystyle I(a_0;a_1,a_3;a_4)\otimes I(a_1;a_2;a_3)}$
\end{tabular}
\end{center}
In addition, we have for the first time a contribution from a term with two complementary polygons (indicated by dashed and dotted lines):
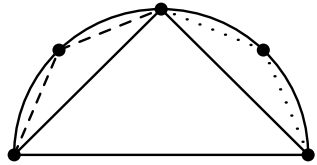
\begin{center}
\begin{tabular}{c}
\begin{fmffile}{W3term8}
\begin{fmfgraph*}(110,70)
\fmfbottom{a0,a4}
\fmfdot{a0}
\fmfdot{a4}
\fmfv{label=$a_0$}{a0}
\fmfv{label=$a_4$}{a4}
\fmf{plain}{a0,a4}
\fmffreeze
\fmf{plain,left,tension=0.2,tag=1}{a0,a4}
\fmfposition
\fmfipath{p[]}
\fmfiset{p1}{vpath1(__a0,__a4)}
\fmfiv{d.sh=circle,d.f=1,d.size=2thick,label=$a_1$}{point length(p1)/4 of p1}
\fmfiv{d.sh=circle,d.f=1,d.size=2thick,label=$a_2$}{point 2length(p1)/4 of p1}
\fmfiv{d.sh=circle,d.f=1,d.size=2thick,label=$a_3$}{point 3length(p1)/4 of p1}
\fmffreeze\fmfposition
\fmfipair{a[]}
\fmfiequ{a1}{point length(p1)/4 of p1}
\fmfiequ{a2}{point 2length(p1)/4 of p1}
\fmfiequ{a3}{point 3length(p1)/4 of p1}
\fmfi{plain}{vloc(__a0) -- a2}
\fmfi{plain}{a2 -- vloc(__a4)}
\fmfi{dashes}{vloc(__a0) -- a1}
\fmfi{dashes}{a1 -- a2}
\fmfi{dots}{a2 -- a3}
\fmfi{dots}{a3 -- vloc(__a4)}
\end{fmfgraph*}
\end{fmffile}\\
${\displaystyle I(a_0;a_2;a_4)\otimes [I(a_0;a_1;a_2)\,I(a_2;a_3;a_4)]}$
\end{tabular}
\end{center}
\end{example}
The worked out case at weight four can be found in ref.~\cite{Duhr:2012fh}.

\paragraph{Shuffle regularisation.} The definition of the coproduct on MPLs given in \eref{eq:coproduct} only holds in the case of generic $a_i$. In order to understand why, let us consider $I(0;1,1;z)$. Looking at Example~\ref{ex:Delta2}, we see that we have a term
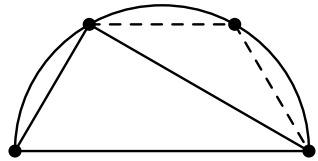
\begin{center}
\begin{tabular}{c}
\begin{fmffile}{NonGen}
\begin{fmfgraph*}(110,70)
\fmfbottom{a0,a3}
\fmfdot{a0}
\fmfdot{a3}
\fmfv{label=$0$}{a0}
\fmfv{label=$z$}{a3}
\fmf{plain}{a0,a3}
\fmffreeze
\fmf{plain,left,tension=0.2,tag=1}{a0,a3}
\fmfposition
\fmfipath{p[]}
\fmfiset{p1}{vpath1(__a0,__a3)}
\fmfiv{d.sh=circle,d.f=1,d.size=2thick,label=$1$}{point length(p1)/3 of p1}
\fmfiv{d.sh=circle,d.f=1,d.size=2thick,label=$1$}{point 2length(p1)/3 of p1}
\fmffreeze\fmfposition
\fmfipair{a[]}
\fmfiequ{a1}{point length(p1)/3 of p1}
\fmfiequ{a2}{point 2length(p1)/3 of p1}
\fmfi{plain}{vloc(__a0) -- a1}
\fmfi{plain}{a1 -- vloc(__a3)}
\fmfi{dashes}{a1 -- a2}
\fmfi{dashes}{a2 -- vloc(__a3)}
\end{fmfgraph*}
\end{fmffile}
 \\
${\displaystyle I(0;1;z)\otimes I(1;1;z)}$  \\
\end{tabular}
\end{center}
The MPL in the second factor of this term in the coproduct is divergent! Indeed, passing to the $G$-notation, we have
$I(1;1;z) = G(1;z)-G(1;1)$, and we have seen in \sref{duhr_sec:MPLs} that $G(a_1,\ldots,a_n;z)$ is divergent whenever $z=a_1$. In other words, the coproduct~\eqref{eq:coproduct} does not make sense for $I(0;1,1;z)$, because it contains divergent quantities (even though the original function $I(0;1,1;z)$ is well-definite).

The idea is now to replace the MPLs that appear inside the coproduct by suitably regularised versions of MPLs~\cite{Goncharov:2001,Goncharov:2005sla}. There are different ways one can define the regularised versions of MPLs. In the following we present the so-called \emph{shuffle regularisation}. The idea is similar to what we did in order to define the regularised shuffle relations: We formally keep all the divergent quantities $G(z,a_2,\ldots,a_n;z)$ in a first step. Then, we proceed in a way similar to what we did in Example~\ref{ex:shuffle_reg}, and we use the shuffle algebra to express all the divergent MPLs in terms of convergent ones, except for MPLs of the form $G(z,\ldots,z;z)$. We then define the \emph{shuffle-regularised version of the MPLs} (denoted by $G^{\textrm{reg}}$ and $I^{\textrm{reg}}$ in the following) by putting all the divergent quantities $G(z,\ldots,z;z)$ to zero.
\begin{example}
\begin{romanlist}
\item Consider the divergent quantity $G(z,\ldots,z;z)$. By definition, its shuffle-regularised value is zero, $G^{\textrm{reg}}(z,\ldots,z;z)=0$.
\item Next consider the divergent quantity $G(z,a;z)$, $a\neq z$. We can write
\beq
G(z,a;z) = G(z;z)\,G(a;z)- G(a,z;z)\,.
\eeq
The regularised value is then
\beq
G^{\textrm{reg}}(z,a;z) = -G(a,z;z)\,.
\eeq
\item Finally, consider the divergent quantity $G(z,z,a;z)$, $a\neq z$. We can write
\beq\bsp
G&(z,z,a;z) = G(z,z;z)\,G(a;z)- G(z,a,z;z)-G(a,z,z;z) \\
&\,= G(z,z;z)\,G(a;z)- [G(z;z)G(a,z;z)-2G(a,z,z;z)]\\
&\,\qquad-G(a,z,z;z)\\
&\,= G(z,z;z)G(a,z)-G(z;z)G(a,z;z) + G(a,z,z;z)\,.
\esp\eeq
The regularised version is then
\beq
G^{\textrm{reg}}(z,z,a;z) = G(a,z,z;z)\,.
\eeq
\end{romanlist}
\end{example}
The coproduct in the non-generic case is now defined by replacing $I$ by $I^{\textrm{reg}}$ everywhere in the right-hand side in \eref{eq:coproduct}. One might wonder why this prescription works (e.g., why it does not spoil any of the other defining conditions of the coproduct). The reason for this is that the regularised versions satisfy the same algebraic properties as the unregularised MPLs. In particular, it is easy to see that the unregularised MPLs agree with the regularised ones whenever they converge. Moreover, the regularisation procedure preserves the multiplication, 
\beq
[G(\vec a;z)\,G(\vec b;z)]^{\textrm{reg}} = G^{\textrm{reg}}(\vec a;z)\,G^{\textrm{reg}}(\vec b;z)\,.
\eeq
\begin{example}
Let us check this last property explicitly on some example:
\beq\bsp
[G&(z,a;z)\,G(b;z)]^{\textrm{reg}} = [G(z,a,b;z)+G(z,b,a;z)+G(b,z,a;z)]^{\textrm{reg}} \\
&\,=G^{\textrm{reg}}(z,a,b;z)+G^{\textrm{reg}}(z,b,a;z)+G^{\textrm{reg}}(b,z,a;z)\\
&\,=-G(a,z,b;z) - G(a,b,z;z)-G(b,a,z;z)\\
&\,=-G(a,z;z)\,G(b;z) \\
&\,=G^{\textrm{reg}}(z,a;z)\,G^{\textrm{reg}}(b;z) \,.
\esp\eeq
\end{example}
Since the regularised and the unregularised MPLs have the same algebraic properties, we will often not make the distinction explicitly, and always assume that inside the coproduct all MPLs have been replaced by their shuffle-regularised versions.

\begin{example} We can now give the coproduct of the classical polylogarithms, and we will be able to give a closed formula for the coproduct of $\li{n}(z)$. To motivate this formula, let us first look at some low weight examples:
\begin{romanlist}
\item Let us start by looking at the coproduct of $I(0;1,0;z)=-\li{2}(z)$. Besides the two trivial ones, we have the following two terms
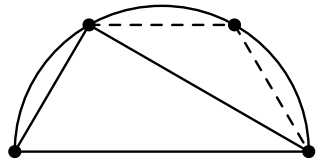
\begin{center}
\begin{tabular}{ccc}
\begin{fmffile}{Li2Term1}
\begin{fmfgraph*}(110,70)
\fmfbottom{a0,a3}
\fmfdot{a0}
\fmfdot{a3}
\fmfv{label=$0$}{a0}
\fmfv{label=$z$}{a3}
\fmf{plain}{a0,a3}
\fmffreeze
\fmf{plain,left,tension=0.2,tag=1}{a0,a3}
\fmfposition
\fmfipath{p[]}
\fmfiset{p1}{vpath1(__a0,__a3)}
\fmfiv{d.sh=circle,d.f=1,d.size=2thick,label=$1$}{point length(p1)/3 of p1}
\fmfiv{d.sh=circle,d.f=1,d.size=2thick,label=$0$}{point 2length(p1)/3 of p1}
\fmffreeze\fmfposition
\fmfipair{a[]}
\fmfiequ{a1}{point length(p1)/3 of p1}
\fmfiequ{a2}{point 2length(p1)/3 of p1}
\fmfi{plain}{vloc(__a0) -- a1}
\fmfi{plain}{a1 -- vloc(__a3)}
\fmfi{dashes}{a1 -- a2}
\fmfi{dashes}{a2 -- vloc(__a3)}
\end{fmfgraph*}
\end{fmffile}
&\phantom{aaaa}
&\begin{fmffile}{Li2Term2}
\begin{fmfgraph*}(110,70)
\fmfbottom{a0,a3}
\fmfdot{a0}
\fmfdot{a3}
\fmfv{label=$0$}{a0}
\fmfv{label=$z$}{a3}
\fmf{plain}{a0,a3}
\fmffreeze
\fmf{plain,left,tension=0.2,tag=1}{a0,a3}
\fmfposition
\fmfipath{p[]}
\fmfiset{p1}{vpath1(__a0,__a3)}
\fmfiv{d.sh=circle,d.f=1,d.size=2thick,label=$1$}{point length(p1)/3 of p1}
\fmfiv{d.sh=circle,d.f=1,d.size=2thick,label=$0$}{point 2length(p1)/3 of p1}
\fmffreeze\fmfposition
\fmfipair{a[]}
\fmfiequ{a1}{point length(p1)/3 of p1}
\fmfiequ{a2}{point 2length(p1)/3 of p1}
\fmfi{plain}{vloc(__a0) -- a2}
\fmfi{plain}{a2 -- vloc(__a3)}
\fmfi{dashes}{vloc(__a0) -- a1}
\fmfi{dashes}{a1 -- a2}
\end{fmfgraph*}
\end{fmffile}
 \\
${\displaystyle I^{\textrm{reg}}(0;1;z)\otimes I^{\textrm{reg}}(1;0;z)}$ && ${\displaystyle I^{\textrm{reg}}(0;0;z)\otimes I^{\textrm{reg}}(0;1;0)}$ \\
${=\displaystyle-\li{1}(z)\otimes \log z}$ && ${\displaystyle =0}$ \\
\end{tabular}
\end{center}
Combining all the pieces, we get
\beq
\Delta(\li{2}(z)) = 1\otimes \li{2}(z) + \li{2}(z)\otimes 1 + \li{1}(z)\otimes \log z\,.
\eeq
\item Next, let us look at the coproduct of $I(0;1,0,0;z)=-\li{3}(z)$. Besides the two trivial ones, we have the following six terms
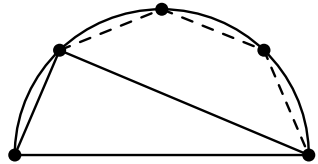
\begin{center}
\begin{tabular}{ccc}
\begin{fmffile}{Li3Term1}
\begin{fmfgraph*}(110,70)
\fmfbottom{a0,a4}
\fmfdot{a0}
\fmfdot{a4}
\fmfv{label=$0$}{a0}
\fmfv{label=$z$}{a4}
\fmf{plain}{a0,a4}
\fmffreeze
\fmf{plain,left,tension=0.2,tag=1}{a0,a4}
\fmfposition
\fmfipath{p[]}
\fmfiset{p1}{vpath1(__a0,__a4)}
\fmfiv{d.sh=circle,d.f=1,d.size=2thick,label=$1$}{point length(p1)/4 of p1}
\fmfiv{d.sh=circle,d.f=1,d.size=2thick,label=$0$}{point 2length(p1)/4 of p1}
\fmfiv{d.sh=circle,d.f=1,d.size=2thick,label=$0$}{point 3length(p1)/4 of p1}
\fmffreeze\fmfposition
\fmfipair{a[]}
\fmfiequ{a1}{point length(p1)/4 of p1}
\fmfiequ{a2}{point 2length(p1)/4 of p1}
\fmfiequ{a3}{point 3length(p1)/4 of p1}
\fmfi{plain}{vloc(__a0) -- a1}
\fmfi{plain}{a1 -- vloc(__a4)}
\fmfi{dashes}{a1 -- a2}
\fmfi{dashes}{a2 -- a3}
\fmfi{dashes}{a3 -- vloc(__a4)}
\end{fmfgraph*}
\end{fmffile}
&\phantom{aaaa}
&\begin{fmffile}{Li3Term2}
\begin{fmfgraph*}(110,70)
\fmfbottom{a0,a4}
\fmfdot{a0}
\fmfdot{a4}
\fmfv{label=$0$}{a0}
\fmfv{label=$z$}{a4}
\fmf{plain}{a0,a4}
\fmffreeze
\fmf{plain,left,tension=0.2,tag=1}{a0,a4}
\fmfposition
\fmfipath{p[]}
\fmfiset{p1}{vpath1(__a0,__a4)}
\fmfiv{d.sh=circle,d.f=1,d.size=2thick,label=$1$}{point length(p1)/4 of p1}
\fmfiv{d.sh=circle,d.f=1,d.size=2thick,label=$0$}{point 2length(p1)/4 of p1}
\fmfiv{d.sh=circle,d.f=1,d.size=2thick,label=$0$}{point 3length(p1)/4 of p1}
\fmffreeze\fmfposition
\fmfipair{a[]}
\fmfiequ{a1}{point length(p1)/4 of p1}
\fmfiequ{a2}{point 2length(p1)/4 of p1}
\fmfiequ{a3}{point 3length(p1)/4 of p1}
\fmfi{dashes}{vloc(__a0) -- a1}
\fmfi{dashes}{a1 -- a2}
\fmfi{dashes}{a2 -- a3}
\fmfi{plain}{vloc(__a0) -- a3}
\fmfi{plain}{a3 -- vloc(__a4)}
\end{fmfgraph*}
\end{fmffile}
 \\
${\displaystyle I^{\textrm{reg}}(0;1;z)\otimes I^{\textrm{reg}}(1;0,0;z)}$ && ${\displaystyle I^{\textrm{reg}}(0;0;z)\otimes  I^{\textrm{reg}}(0;1,0;0)}$ \\
${\displaystyle =\li{1}(z)\otimes \frac{\log^2 z}{2}}$ && ${\displaystyle =0}$ \\
\end{tabular}
\end{center}
\begin{center}
\begin{tabular}{ccc}
\begin{fmffile}{Li3Term3}
\begin{fmfgraph*}(110,70)
\fmfbottom{a0,a4}
\fmfdot{a0}
\fmfdot{a4}
\fmfv{label=$0$}{a0}
\fmfv{label=$z$}{a4}
\fmf{plain}{a0,a4}
\fmffreeze
\fmf{plain,left,tension=0.2,tag=1}{a0,a4}
\fmfposition
\fmfipath{p[]}
\fmfiset{p1}{vpath1(__a0,__a4)}
\fmfiv{d.sh=circle,d.f=1,d.size=2thick,label=$1$}{point length(p1)/4 of p1}
\fmfiv{d.sh=circle,d.f=1,d.size=2thick,label=$0$}{point 2length(p1)/4 of p1}
\fmfiv{d.sh=circle,d.f=1,d.size=2thick,label=$0$}{point 3length(p1)/4 of p1}
\fmffreeze\fmfposition
\fmfipair{a[]}
\fmfiequ{a1}{point length(p1)/4 of p1}
\fmfiequ{a2}{point 2length(p1)/4 of p1}
\fmfiequ{a3}{point 3length(p1)/4 of p1}
\fmfi{plain}{vloc(__a0) -- a1}
\fmfi{plain}{a1 -- a2}
\fmfi{plain}{a2 -- vloc(__a4)}
\fmfi{dashes}{a2 -- a3}
\fmfi{dashes}{a3 -- vloc(__a4)}
\end{fmfgraph*}
\end{fmffile}
&\phantom{aaaa}
&\begin{fmffile}{Li3Term4}
\begin{fmfgraph*}(110,70)
\fmfbottom{a0,a4}
\fmfdot{a0}
\fmfdot{a4}
\fmfv{label=$0$}{a0}
\fmfv{label=$z$}{a4}
\fmf{plain}{a0,a4}
\fmffreeze
\fmf{plain,left,tension=0.2,tag=1}{a0,a4}
\fmfposition
\fmfipath{p[]}
\fmfiset{p1}{vpath1(__a0,__a4)}
\fmfiv{d.sh=circle,d.f=1,d.size=2thick,label=$1$}{point length(p1)/4 of p1}
\fmfiv{d.sh=circle,d.f=1,d.size=2thick,label=$0$}{point 2length(p1)/4 of p1}
\fmfiv{d.sh=circle,d.f=1,d.size=2thick,label=$0$}{point 3length(p1)/4 of p1}
\fmffreeze\fmfposition
\fmfipair{a[]}
\fmfiequ{a1}{point length(p1)/4 of p1}
\fmfiequ{a2}{point 2length(p1)/4 of p1}
\fmfiequ{a3}{point 3length(p1)/4 of p1}
\fmfi{dashes}{vloc(__a0) -- a1}
\fmfi{dashes}{a1 -- a2}
\fmfi{plain}{vloc(__a0) -- a2}
\fmfi{plain}{a2 -- a3}
\fmfi{plain}{a3 -- vloc(__a4)}
\end{fmfgraph*}
\end{fmffile}
 \\
${\displaystyle I^{\textrm{reg}}(0;1,0;z)\otimes I^{\textrm{reg}}(0;0;z)}$ && ${\displaystyle I^{\textrm{reg}}(0;0,0;z)\otimes  I^{\textrm{reg}}(0;1;0)}$ \\
${\displaystyle =\li{2}(z)\otimes \log z}$ && ${\displaystyle =0}$ \\
\end{tabular}
\end{center}
\begin{center}
\begin{tabular}{ccc}
\begin{fmffile}{Li3Term5}
\begin{fmfgraph*}(110,70)
\fmfbottom{a0,a4}
\fmfdot{a0}
\fmfdot{a4}
\fmfv{label=$0$}{a0}
\fmfv{label=$z$}{a4}
\fmf{plain}{a0,a4}
\fmffreeze
\fmf{plain,left,tension=0.2,tag=1}{a0,a4}
\fmfposition
\fmfipath{p[]}
\fmfiset{p1}{vpath1(__a0,__a4)}
\fmfiv{d.sh=circle,d.f=1,d.size=2thick,label=$1$}{point length(p1)/4 of p1}
\fmfiv{d.sh=circle,d.f=1,d.size=2thick,label=$0$}{point 2length(p1)/4 of p1}
\fmfiv{d.sh=circle,d.f=1,d.size=2thick,label=$0$}{point 3length(p1)/4 of p1}
\fmffreeze\fmfposition
\fmfipair{a[]}
\fmfiequ{a1}{point length(p1)/4 of p1}
\fmfiequ{a2}{point 2length(p1)/4 of p1}
\fmfiequ{a3}{point 3length(p1)/4 of p1}
\fmfi{plain}{vloc(__a0) -- a1}
\fmfi{plain}{a1 -- a3}
\fmfi{plain}{a3 -- vloc(__a4)}
\fmfi{dashes}{a1 -- a2}
\fmfi{dashes}{a2 -- a3}
\end{fmfgraph*}
\end{fmffile}
&\phantom{aaaa}
&\begin{fmffile}{Li3Term6}
\begin{fmfgraph*}(110,70)
\fmfbottom{a0,a4}
\fmfdot{a0}
\fmfdot{a4}
\fmfv{label=$0$}{a0}
\fmfv{label=$z$}{a4}
\fmf{plain}{a0,a4}
\fmffreeze
\fmf{plain,left,tension=0.2,tag=1}{a0,a4}
\fmfposition
\fmfipath{p[]}
\fmfiset{p1}{vpath1(__a0,__a4)}
\fmfiv{d.sh=circle,d.f=1,d.size=2thick,label=$1$}{point length(p1)/4 of p1}
\fmfiv{d.sh=circle,d.f=1,d.size=2thick,label=$0$}{point 2length(p1)/4 of p1}
\fmfiv{d.sh=circle,d.f=1,d.size=2thick,label=$0$}{point 3length(p1)/4 of p1}
\fmffreeze\fmfposition
\fmfipair{a[]}
\fmfiequ{a1}{point length(p1)/4 of p1}
\fmfiequ{a2}{point 2length(p1)/4 of p1}
\fmfiequ{a3}{point 3length(p1)/4 of p1}
\fmfi{dashes}{vloc(__a0) -- a1}
\fmfi{dashes}{a1 -- a2}
\fmfi{dots}{a2 -- a3}
\fmfi{dots}{a3 -- vloc(__a4)}
\fmfi{plain}{vloc(__a0) -- a2}
\fmfi{plain}{a2 -- vloc(__a4)}
\end{fmfgraph*}
\end{fmffile}
 \\
${\displaystyle I^{\textrm{reg}}(0;1,0;z)\otimes I^{\textrm{reg}}(1;0;0)}$ && ${\displaystyle I^{\textrm{reg}}(0;0;z)}$ \\
${\displaystyle }$ && ${\displaystyle \otimes  [I^{\textrm{reg}}(0;1;0)\,I^{\textrm{reg}}(0;0;z)]}$ \\
${\displaystyle =0}$ && ${\displaystyle =0}$ \\
\end{tabular}
\end{center}
Putting all the terms together, we find
\beq\bsp
\Delta(\li{3}(z)) &\,= 1\otimes\li{3}(z) + \li{3}(z)\otimes1\\
&\,+\li{2}(z)\otimes \log z +\li{1}(z)\otimes\frac{\log^2z}{2}\,.
\esp\eeq
\item From the previous examples, it is easy to discern a pattern for all classical polylogarithms: the only non-vanishing contributions are those where the inscribed polygon contains the vertex labeled by `1' and where there is exactly one complementary polygon. These contribution are easy to evaluate, and we find
\beq\label{eq:Delta_Li_n}
\Delta(\li{n}(z)) = 1\otimes\li{n}(z) + \sum_{k=0}^{n-1}\li{n-k}(z)\otimes \frac{\log^kz}{k!}\,.
\eeq
\end{romanlist}
\end{example}

\paragraph{Inclusion of even zeta values.} We know now how to compute the coproduct of arbitrary MPLs. In particular, this implies that we also know how to compute the coproduct of MZVs, by writing them as MPLs evaluated at $1$. For example, letting $z=1$ in \eref{eq:Delta_Li_n}, we see that ordinary zeta values are primitive,
\beq\label{eq:Delta_zeta_n}
\Delta(\zeta_n) = \zeta_n\otimes 1+1\otimes\zeta_n\,.
\eeq
This, however, is problematic for even zeta values. Indeed, we find
\beq
\Delta(\zeta_4)=\frac{2}{5}\,\Delta(\zeta_2)^2 = \zeta_4\otimes 1+1\otimes \zeta_4+\frac{4}{5}\zeta_2\otimes\zeta_2\,,
\eeq
i.e., we have obtained a contradiction with \eref{eq:Delta_zeta_n} for $n=4$. It is easy to see that a similar problem arises for $i\pi$ ($=\log(-1)$). 

One way to resolve the contradiction is to work modulo $\zeta_2$ and $i\pi$ (i.e., `$\zeta_2=i\pi=0$'), and indeed, $\mathcal{A}$ is, strictly speaking, \emph{not} a Hopf algebra. If we define $\mathcal{H}$ to be the algebra $\mathcal{A}$ modulo $i\pi$, then $\mathcal{H}$ is a Hopf algebra with the coproduct given in \eref{eq:coproduct}~\cite{Goncharov:2005sla}. It is clear, however, that this situation is not satisfactory from a practical point of view.

In the following we discuss how to remedy this problem without having to work modulo $i\pi$, and we follow very closely ideas introduced by Brown in ref.~\cite{Brown:2011ik} in the context of MZVs. First, we note that we can trivially write
\beq
\mathcal{A} = \mathbb{Q}[i\pi]\otimes \mathcal{H}\,,
\eeq
where $\mathbb{Q}[i\pi]$ denotes the ring of polynomials in $i\pi$ with rational coefficients. The meaning of this is simply that by passing from $\mathcal{A}$ to $\mathcal{H}$, we have removed all powers of $i\pi$, and we can compensate for this by allowing the coefficients in front of elements of $\mathcal{H}$ to be polynomials in $i\pi$ rather than just rational numbers. Next, we define the coproduct\footnote{Strictly speaking, $\Delta$ is no longer a coproduct, but a coaction, and $\mathcal{A}$ is a comodule rather than a Hopf algebra. Since this distinction is only purely technical for our purposes, we will continue to call $\mathcal{A}$ a `Hopf algebra', keeping in mind that we need this special treatment of $i\pi$.} to be a map $\Delta:\mathcal{A}\to \mathcal{A}\otimes \mathcal{H}$ in such a way that it acts by \eref{eq:coproduct} on elements of $\mathcal{H}$, while on $i\pi$ it acts by~\cite{Brown:2011ik,Duhr:2012fh}
\beq\label{eq:DeltaIPi}
\Delta(i\pi) = i\pi\otimes 1\,.
\eeq
Practically speaking, this means that we have to put $i\pi$ to zero everywhere but in the left-most factor of the coproduct.
This obviously resolves the aforementioned contradiction in a trivial way, because
\beq
\Delta(\zeta_4) = \zeta_4\otimes 1 = \frac{2}{5}\,\zeta_2^2\otimes1 = \frac{2}{5}\,\Delta(\zeta_2)^2\,.
\eeq

This completes our review of the Hopf algebra of MPLs. We note that the fact that the coproduct maps $\mathcal{A}$ to $\mathcal{A}\otimes\mathcal{H}$ introduces an `asymmetry' between the left and right factors, and we may ask what the meaning of this asymmetry is. This will be discussed in the rest of this section.

Let us start by analysing the rightmost factor of the coproduct. It turns out that this factor encodes the behaviour of the functions under differentiation. More precisely, we have
\beq\label{eq:Delta_diff}
\Delta\left(\frac{\partial}{\partial z}F\right) = \left(\textrm{id}\otimes\frac{\partial}{\partial z}\right)\Delta(F)\,,
\eeq
i.e., derivatives only act in the rightmost factor of the coproduct. Note that this gives a convenient way to compute the derivatives of MPLs with respect to arbitrary variables. Indeed, if $F$ denotes a function of weight $n$, we have
\beq\label{eq:derG}
\frac{\partial}{\partial z}F = \mu\left(\textrm{id}\otimes\frac{\partial}{\partial z}\right)\Delta_{n-1,1}(F)\,,
\eeq
where $\mu(a\otimes b)=ab$ denotes multiplication.
\begin{example}
 Let us illustrate \eref{eq:Delta_diff} on the simple example of the dilogarithm. The left-hand side of  \eref{eq:Delta_diff} gives
\beq
\Delta\left(\frac{\partial}{\partial z}\li{2}(z)\right) = \Delta\left(\frac{1}{ z}\li{1}(z)\right)=\frac{1}{z}\,(1\otimes\li{1}(z)+\li{1}(z)\otimes 1)\,.
\eeq
The right-hand side gives
\beq\bsp
\Big(\textrm{id}&\otimes\frac{\partial}{\partial z}\Big)\Delta(\li{2}(z))\\
&\,=1\otimes\frac{\partial}{\partial z}\li{2}(z)+\li{2}(z)\otimes \frac{\partial}{\partial z}1+\li{1}(z)\otimes\frac{\partial}{\partial z}\log z\\
&\,=1\otimes\frac{\li{1}(z)}{z}+\li{1}(z)\otimes\frac{1}{z}\\
&\,=\frac{1}{z}\,(1\otimes\li{1}(z)+\li{1}(z)\otimes 1)\,,
\esp\eeq
and we indeed obtain the same answer.
\end{example}

\begin{example}
Assume that you want to compute the derivative of $G(1,1+y;z)$ with respect to $y$. Using \eref{eq:derG}, we obtain
\beq\bsp
\frac{\partial}{\partial y}&G(1,1+y;z) = \mu\left(\textrm{id}\otimes\frac{\partial}{\partial z}\right)\Delta_{1,1}(G(1,1+y;z))\\
&\,=G(1;z)\, \frac{\partial}{\partial y}G(1+y;1)-G(1+y;z)\,\frac{\partial}{\partial y} G(1;1+y)\\
&\,+G(1+y;z)\,\frac{\partial}{\partial y} G(1;z)\,.
\esp\eeq
The remaining derivatives are just derivatives of ordinary logarithms,
\begin{eqnarray}
\frac{\partial}{\partial y}G(1+y;1) &=& \frac{\partial}{\partial y}\log\left(1-\frac{1}{1+y}\right) = \frac{1}{y\,(1+y)}\,,\\
\frac{\partial}{\partial y} G(1;1+y)&=& \frac{\partial}{\partial y} \log(-y)= \frac{1}{y}\,,\\
\frac{\partial}{\partial y} G(1;z) &=& 0\,.
\end{eqnarray}
Thus,
\beq
\frac{\partial}{\partial y}G(1,1+y;z) = \frac{1}{y\,(1+y)}\,G(1;z)-\frac{1}{y}\,G(1+y;z)\,.
\eeq
\end{example}

Just like the rightmost factor of the coproduct encodes the derivatives of a function, the leftmost factor encodes its discontinuities. Note that this is consistent with the fact that $\Delta(i\pi)=i\pi\otimes 1$. If Disc denotes the operator that takes the discontinuity of a function across some branch cut, then we have
\beq\label{eq:Delta_Disc}
\Delta\left(\textrm{Disc}F\right)=\left(\textrm{Disc}\otimes\textrm{id}\right)\Delta(F)\,.
\eeq 
\begin{example}
We again illustrate this property on the example of the dilogarithm. $\textrm{Li}_2(z)$ has a branch cut
extending from $z=1$ to $z=\infty$, and the discontinuity across the cut is
\beq
\textrm{Disc}\,\li{2}(z) = \li{2}(z+i0)-\li{2}(z-i0) = 2\pi i\,\log z\,.
\eeq
The right-hand side of \eref{eq:Delta_Disc} gives
\beq
\Delta\left(\textrm{Disc}\,\li{2}(z)\right) = 2\pi i\otimes \log z + (2\pi i\,\log z)\otimes 1\,.
\eeq
The left-hand side gives
\beq\bsp
\Big(\textrm{Disc}&\otimes\textrm{id}\Big)\Delta(\li{2}(z)) \\
&\,= \textrm{Disc}\,1\otimes \li{2}(z) + \textrm{Disc}\,\li{2}(z)\otimes1+\textrm{Disc}\,\li{1}(z)\otimes\log z\\
&\,= (2\pi i\,\log z)\otimes 1+2\pi i\otimes \log z\,,
\esp\eeq
where we used the fact that $\textrm{Disc}\,\li{1}(z) = 2\pi i$.
\end{example}

\subsection{The symbol map}
In the previous section we have seen that we can decompose an MPL of weight $n$ into smaller weights by acting with the coproduct, and coassociativity allows us to iterate this decomposition in a unique way. This decomposition will obviously stop at some point, namely when we have decomposed a function of weight $n$ into an $n$-fold tensor product of functions of weight one, i.e., ordinary logarithms. This maximal iteration of the coproduct has a special status, and is often referred to as the \emph{symbol} in the literature~\cite{Chen,Goncharov:2010jf,Brown:2009qja,Goncharov:2005sla,Duhr:2011zq},
\beq
\cS(F) \equiv \Delta_{1,\ldots,1}(F)\,\mod \,i\pi\,,
\eeq
where we also put all $i\pi$ terms to zero. Note that, since all the factors in the symbol are just ordinary logarithms, it is conventional to drop the `log'-signs inside the factors of the tensor product, i.e., we write $a_1\otimes\ldots\otimes a_n$ instead of $\log a_1\otimes\ldots\otimes\log a_n$. The entries $a_i$ in the symbol of $F$ are often referred to as the \emph{alphabet} of $F$.

In the following we give some properties of the symbol map $\cS$, most of which are direct consequences of the corresponding properties of the coproduct. First, it is obvious that $\cS$ is linear. Second, the symbols of a product is mapped to the shuffle of the symbols,
\beq
\cS(F\,G) = \cS(F)\shuffle\cS(G)\,,
\eeq
where $\shuffle$ denotes the shuffle product on tensors, e.g.,
\beq\bsp
(a\otimes b)&\shuffle(c\otimes d) = a\otimes b\otimes c\otimes d + a\otimes c\otimes b\otimes d+c\otimes a\otimes b\otimes d\\
&\, + a\otimes c\otimes d\otimes b + c\otimes a\otimes d\otimes b + c\otimes d\otimes a\otimes b\,.
\esp\eeq
Next, the additivity of the logarithm, $\log(ab)=\log a+\log b$ translates into the property
\begin{eqnarray}
\ldots\otimes(ab)\otimes\ldots &=& \ldots\otimes a\otimes \ldots + \ldots\otimes b\otimes\ldots\,,\\
\ldots\otimes a^n\otimes\ldots &=& n\,(\ldots\otimes a\otimes \ldots)\,,
\end{eqnarray}
and the fact that we work modulo $i\pi$ leads to
\beq
\ldots \otimes \rho\otimes \ldots =0\,,
\eeq 
where $\rho$ is a root of unity, $\rho^n=1$ for some $n$. This last property implies that $\cS$ has a non-trivial kernel. In particular, the kernel contains all MZVs, $\cS(\zeta_{m_1,\ldots,m_k})=0$, but it contains additional non-trivial elements that are not necessarily MZVs, e.g., 
\beq
\cS\left[\li{4}\left(\frac{1}{2}\right)+\frac{1}{24}\log^42\right] = 0\,.
\eeq
A collection of elements in the kernel of the symbol map is given in ref.~\cite{Duhr:2011zq}.

Finally, we may ask if every possible tensor 
\beq
S = \sum_{i_1,\ldots,i_k}c_{i_1,\ldots, i_k}\,a_{i_1}\otimes\ldots \otimes a_{i_k}\,,
\eeq
can be the symbol of some function, i.e., whether we can find a function $F$ such that $\cS(F)=S$. The answer to this question is negative in general, but one can show that such a function $F$ exists if and only if  $S$ satisfies the \emph{integrability condition}~\cite{Chen}, for all $1\le j \le k-1$,
\beq
\sum_{i_1,\ldots,i_k}c_{i_1,\ldots, i_k}\,d\log a_{i_j}\wedge d\log a_{i_{j+1}}\,a_{i_1}\otimes\ldots a_{i_{j-1}}\otimes a_{i_{j+2}} \otimes a_{i_k} = 0\,,
\eeq
where $\wedge$ denotes the usual wedge product on differential forms.

\subsection{Functional equations of MPLs}
In this section we discuss our main application of the Hopf algebra of MPLs: the derivation of functional equations for MPLs. The idea is simple: Assume we are trying to proof an identity $F=G$, where $F$ and $G$ are expressions of weight $n$. We can decompose this expression into lower weights using the coproduct, and prove a sequence of simpler identities instead, which only involve simpler functions (where `simpler' means `lower weights'), for which we may assume that all identities are known. We will illustrate this procedure on some examples in the following. Note that in practise the expressions can soon become rather big. The examples in the following are chosen because they are simple enough so that all the manipulations can be carried out on a piece of paper.

\begin{example}
Throughout this section we assume that $x$ is a real positive variable to which we assign a small positive imaginary part.
We proceed recursively in the weight to build up the inversion relations for classical polylogarithms.
 
For the classical polylogarithm of weight one, the inversion relation is easy to obtain, 
\beq\bsp\label{eq:inversion_Li1}
\li{1}\left({1\over x}\right) &\,= -\log\left(1-{1\over x}\right) = -\log(1-x) +\log(-x)\\
&\, = -\log(1-x) + \log x - i\pi\,.
\esp\eeq
In order to obtain the inversion relation for weight two, we act with $\Delta_{1,1}$ on $\li{2}(1/x)$ and insert the inversion relation for $\li{1}(1/x)$,
\beq\bsp
\Delta_{1,1}\left[\li{2}\left({1\over x}\right)\right] &\,= -\log\left(1-{1\over x}\right) \otimes \log\left({1\over x}\right) \\
&\, = \log(1-x)\otimes\log x - \log x\otimes \log x +i\pi\otimes \log x\\
 &\,=\Delta_{1,1}\Big[-\li{2}(x) -{1\over 2}\log^2x + i\pi\log x\Big]\,.
\esp\eeq
Note that in the last step \eref{eq:DeltaIPi} played a crucial role.
We conclude that the arguments in the left and right-hand sides are equal modulo primitive elements of weight two. We thus make the ansatz,
\beq
\li{2}\left({1\over x}\right) = -\li{2}(x) -{1\over 2}\log^2x + i\pi\log x + c\,\zeta_2\,,
\eeq
for some rational number $c$. Specializing to $x=1$, we immediately obtain $c=2$, which is indeed the correct inversion relation for $\li2$. 

At weight three, we act with $\Delta_{1,1,1}$ on $\li{3}(1/x)$ and we obtain
\beq\bsp\label{eq:temp}
\Delta_{1,1,1}\left[\li{3}\left({1\over x}\right)\right] &\,= -\log\left(1-{1\over x}\right) \otimes \log\left({1\over x}\right)\otimes \log\left({1\over x}\right) \\
&\,= -\log(1-x)\otimes\log x\otimes\log x + \log x\otimes\log x\otimes\log x\\
&\, - i\pi\otimes\log x\otimes \log x\\
&\, = \Delta_{1,1,1}\Big[\li{3}(x) +{1\over 6}\log^3x - {i\pi\over 2}\log^2 x\Big]\,.
\esp\eeq
Eq.~\eqref{eq:temp} is not yet the correct inversion relation for $\li{3}$.
After subtracting the terms we have found in Eq.~\eqref{eq:temp}, we look at the image of the difference under $\Delta_{2,1}$ or $\Delta_{1,2}$. For example, we obtain
\beq\bsp
\Delta&_{1,2}\Bigg[\li{3}\left({1\over x}\right) - \Big(\li{3}(x) +{1\over 6}\log^3x - {i\pi\over 2}\log^2 x\Big)\Bigg] \\ 
&= -{1\over 2}\log\left(1-{1\over x}\right)\otimes\log^2\left({1\over x}\right) + {1\over2}\log(1-x)\otimes\log^2 x\\
&\,-{1\over2}\log x\otimes \log^2x + {i\pi\over 2}\otimes \log^2x\\
& = 0\,.
\esp\eeq
We see that acting with $\Delta_{1,2}$ does not provide any new information. This is not surprising, as the missing terms are of the form $\zeta_2\log x$, and $\Delta_{1,2}(\zeta_2\log x) = 0$. Acting with $\Delta_{2,1}$ and using the inversion relation for $\li{2}$, we obtain new non-trivial information,
\beq\bsp
\Delta&_{2,1}\Bigg[\li{3}\left({1\over x}\right) - \Big(\li{3}(x) +{1\over 6}\log^3x - {i\pi\over 2}\log^2 x\Big)\Bigg] \\ 
&= \li{2}\left({1\over x}\right)\otimes\log\left({1\over x}\right) -\li{2}(x)\otimes\log x-{1\over2}\log^2 x\otimes \log x \\
&\,\quad+ (i\pi\log x)\otimes \log x\\
& = -\Big[-\li{2}(x)-{1\over2}\log^2x + i\pi\log x + 2\zeta_2\Big]\otimes \log x\\
&\,\quad -\li{2}(x)\otimes\log x-{1\over2}\log^2 x\otimes \log x + (i\pi\log x)\otimes \log x\\
&= -2\zeta_2\otimes\log x\\
& = \Delta_{2,1}\Big(-2\zeta_2\log x\Big)\,.
\esp\eeq
Thus,
\beq
\li{3}\left({1\over x}\right) = \li{3}(x) +{1\over 6}\log^3x - {i\pi\over 2}\log^2 x - 2\zeta_2\log x + \alpha\zeta_3 + \beta\,i\pi^3\,.
\eeq
Specializing to $x=1$ gives $\alpha = \beta = 0$, which is indeed the correct inversion relation for $\li{3}$. 
Proceeding in exactly the same way, we can now derive the inversion relations for all the classical polylogarithms.
\end{example}


\section{Applications to loop amplitudes}
\label{duhr_sec:amps}

\subsection{MPLs and Feynman integrals}

In this section we give examples of how the concepts introduced in the previous sections apply to loop integrals. Since the area of potential applications of the Hopf algebraic techniques are very wide, we do by now means intent to be exhaustive, but we only try to give a flavour of what kind of applications are possible. Note that we will restrict ourselves to classes of loop integrals that can be expressed in terms of MPLs, keeping in mind that this may not always be possible. 

The first question one may ask is how the loop integrals themselves fit into the algebraic picture of the previous section. It is well-known that Feynman integrals have discontinuities, and the locations of the branch cuts are solutions of the so-called Landau equations~\cite{Landau}. For example, in the particular case of massless propagators all branch cuts start at points where a Mandelstam invariant becomes zero or infinite. Thus, the position of the branch points of Feynman integrals (seen as functions of Lorentz invariant scalar products and masses) are not arbitrary, but dictated by unitarity. In \eref{eq:Delta_Disc} we have seen that discontinuities are captured by the leftmost factor of the coproduct of a function. It then follows that the leftmost factor in the coproduct can only have discontinuities which are compatible with the Landau equations! This condition, known as the \emph{first entry condition}, puts strong constraints on the analytic expressions for Feynman integrals. In particular, for massless propagators this implies that the first entry in the symbol of such an integral can only be a Mandelstam invariant~\cite{Gaiotto:2011dt}.

The next question is whether one can make any kind of generic statements about the weight of loop amplitudes (at least in the case where they can be expressed in terms of MPLs). Currently, only very few theorems are known for specific classes of loop integrals~\cite{Brown:2009rc,Brown:2012ah}, but there are conjectures about the weight of generic Feynman integrals in four dimensions:
\begin{conjecture}\label{eq:MaxTrans} 
In $D=4-2\eps$ dimensions, the Laurent coefficient of $\eps^k$ of an $L$-loop amplitude contains terms of weight at most $2L+k$.
\end{conjecture}
Note that the conjecture only gives an upper bound for the weight. In general, a loop amplitude will contain all weights up to the bound given by the conjecture. In special quantum field theories, like for example the $\mathcal{N}=4$ Super Yang-Mills (SYM) theory, this bound is expected to be saturated, i.e., in the coefficient of $\eps^k$ of an $L$-loop amplitudes has  \emph{exactly} weight $2L+k$. We emphasise that these conjectures only make sense if the amplitude can be expressed in terms of MPLs in the first place.

The previous conjecture only allows one to set an upper bound on the weight of Feynman integrals, and so one may ask  whether there is a more refined version of it, e.g., is it possible to find `building blocks' that are of \emph{uniform weight} (i.e., where all the terms in a given Laurent coefficient of have the same weight) and out of which the amplitude can be constructed. It turns out that this is indeed the case, but in order to formulate the corresponding conjecture we need to introduce a few more concepts.

Let us start by defining more precisely what we mean by `building blocks' of the loop amplitude. It is clear that the amplitude is a linear combination of scalar integrals of the type~\eqref{eq:feyn_int}, and let us concentrate on a specific subset of of these integrals that have the same set of propagators (and we may interpret numerators as propagators raised to negative powers). These integrals in general differ only by the powers $\nu_i$ of the propagators\footnote{We allow for some propagators to be absent, i.e., they are raised to the power zero.}. Such a family of loop integrals is referred to as a \emph{topology}. More precisely, we can define a topology to be a family of scalar integrals differing only by the exponents $\nu_i\in\mathbb{Z}$ such that every scalar product between two loop momenta, $k_i\cdot k_j$, or between a loop and an external momentum, $k_i\cdot p_j$, can be written as a linear combination of propagators. This latter condition ensures that all the numerator factors can be interpreted as propagators raised to non-positive powers.

\begin{example}
Let us consider the three integrals
\begin{eqnarray}
I_1 &=& e^{\gamma_E\eps}\int\frac{d^Dk}{i\pi^{D/2}}\frac{1}{k^2\,(k+p)^2}\,,\\
I_2 &=& e^{\gamma_E\eps}\int\frac{d^Dk}{i\pi^{D/2}}\frac{k\cdot p_1}{k^2\,(k+p_1)^2\,(k+p_1+p_2)^2}\,,\\
I_3 &=& e^{\gamma_E\eps}\int\frac{d^Dk}{i\pi^{D/2}}\frac{k^2+2k\cdot p_2}{k^2\,(k+p_1)^2\,(k+p_1+p_2)^2\,(k-p_4)^2}\,.
\end{eqnarray}
All three integrals can in fact be embedded into the same \emph{box topology},
\beq\bsp
&\textrm{Box}(\nu_1,\nu_2,\nu_3,\nu_4)\\
&\, = e^{\gamma_E\eps}\int\frac{d^Dk}{i\pi^{D/2}}\frac{1}{[k^2]^{\nu_1}\,[(k+p_1)^2]^{\nu_2}\,[(k+p_1+p_2)^2]^{\nu_3}\,[(k-p_4)^2]^{\nu_4}}\,.
\esp\eeq
Indeed, the two-point integral can be written as $I_1=\textrm{Box}(1,1,0,0)$. For the three-point integral $I_2$, we can rewrite the numerator in terms of propagators
\beq
k\cdot p_1 = \frac{1}{2}[(k+p_1)^2 - k^2-p_1^2]\,,
\eeq
and so we get
\beq
I_2=\frac{1}{2}\textrm{Box}(1,0,1,0) - \frac{1}{2}\textrm{Box}(0,1,1,0) - \frac{1}{2}p_1^2\,\textrm{Box}(1,1,1,0)\,.
\eeq
Similarly, the numerator of the four-point integrals $I_3$ can written as
\beq
k^2+2k\cdot p_2 = (k+p_1+p_2)^2- (k+p_1)^2 + k^2+p_1^2 - (p_1+p_2)^2\,,
\eeq
and so
\beq\bsp
I_3 = &\, \textrm{Box}(1,1,0,1) - \textrm{Box}(1,0,1,1) \\
&\,+ \textrm{Box}(0,1,1,1) +[p_1^2 - (p_1+p_2)^2]\,\textrm{Box}(1,1,1,1)\,.
\esp\eeq
\end{example}
Let us now consider a topology, and let us see it as a function of the powers $\nu_i$ of the propagators. If we work in dimensional regularisation, then the integral of a total derivative vanishes, i.e., we can write
\beq
0 = \int \left(\prod_{i=1}^L\,d^Dk_i\right)\,\frac{\partial}{\partial k_j^\mu}\cdot v^\mu\Big(\ldots\Big)\,,
\eeq
where $v^\mu\in\{k_1^\mu,\ldots,k_L^\mu,p_1^\mu,\ldots,p_{E-1}^\mu\}$ can be either a loop momentum or an external momentum. If we act with the derivative on the propagators, we will essentially only shift the powers of the propagators, and so we will eventually arrive at a set of linear recursion relations in the powers $\nu_i$ of the propagators, known as \emph{integration-by-parts identities (IBPs)}~\cite{Chetyrkin:1981qh}. These recursion relations can be solved algorithmically, e.g., using Laporta's algorithm~\cite{Laporta:1996mq,Laporta:2001dd,Anastasiou:2004vj,Smirnov:2008iw,Smirnov:2013dia,Smirnov:2014hma,vonManteuffel:2012np,Lee:2012cn}. Since the recursion is linear, we can express all the integrals in the topology in terms of a basis of the solution space of the linear system. Such a basis is referred to as a set of \emph{master integrals}, and they are the `building blocks' we were looking for.

\begin{example} Consider the topology defined by
\beq
\textrm{Bub}(\nu_1,\nu_2) =  e^{\gamma_E\eps}\int\frac{d^Dk}{i\pi^{D/2}}\frac{1}{[k^2]^{\nu_1}\,[(k+p)^2]^{\nu_2}}\,.
\eeq
We can write down two IBP relations,
\begin{eqnarray}
0 & = & (D-2\nu_1-\nu_2)\,\textrm{Bub}(\nu_1,\nu_2)-\nu_2\,\textrm{Bub}(\nu_1-1,\nu_2+1)\\
&&+\nu_2\,p^2\,\textrm{Bub}(\nu_1,\nu_2+1)\,,\nonumber\\
0 & = & (\nu_1-\nu_2)\,\textrm{Bub}(\nu_1,\nu_2) + \nu_1\,p^2\,\textrm{Bub}(\nu_1+1,\nu_2)\\
&& - \nu_2\,p^2\,\textrm{Bub}(\nu_1,\nu_2+1)  - \nu_1\,\textrm{Bub}(\nu_1+1,\nu_2-1) \nonumber\\
&& + \nu_2\,\textrm{Bub}(\nu_1-1,\nu_2+1)\,.\nonumber
\end{eqnarray}
If we solve the recursion, we see that there is only one single master integral, $\textrm{Bub}(1,1)$, which is the usual one-loop bubble integral. All other integrals in the topology can be expressed in terms of this integral, e.g.,
\beq
\textrm{Bub}(2,3) = -\frac{(D-8)(D-5)(D-4)}{2\,(p^2)^3}\,\textrm{Bub}(1,1)\,.
\eeq
\end{example}
While it can be proven that the number of master integrals is always finite~\cite{Smirnov:2010hn}, and their number can be predicted from the topology~\cite{Lee:2013hzt}, there are of course different ways of choosing the set of master integrals. Recently, it was conjectured there is a distinguished set of master integrals for every topology~\cite{Henn:2013pwa}:
\begin{conjecture}
For every topology there is a set of uniformly transcendental master integrals with unit leading singularity.
\end{conjecture}
While the conjecture asserts the existence of this set of master integrals, no generic algorithm is known to find this basis (some partial results to find the basis exist, see, e.g., ref.~\cite{Henn:2013pwa,Gehrmann:2014bfa,Caron-Huot:2014lda,Argeri:2014qva}). The main advantage of having this uniformly transcendental basis is that, once this basis is known, the master integrals can easily be computed via differential equations. Indeed, it is well-known that master integrals satisfy coupled systems of first-order differential equations~\cite{Kotikov:1991hm,Kotikov:1991pm,Gehrmann:1999as}. If the basis of master integrals is uniformly transcendental, then the differential equations decouple order-by-order in $\eps$ and can easily be solved~\cite{Henn:2013pwa}. In fact, the differential equations satisfied by uniformly transcendental master integrals are special instances/generalisations of the so-called Knizhnik-Zalmolodchikov equation. Although this topic is tightly connected to iterated integrals and MPLs, this topic would lead us to far, and we will therefore not cover it in this set of lectures.

\subsection{Simplification of analytic results}
Probably the most obvious application to loop amplitudes of the ideas presented in these lectures is the simplification of the sometimes large and complicated results that arise from these computations. For example, analytic results for two-loop multi-scale Feynman integrals and scattering amplitudes may involve combinations of several thousands of MPLs, and so the question whether a given a expression can be rewritten in terms of `simpler' functions and/or in a more compact form is highly relevant. The relation between the `complicated' and the `simple' results may be seen as one big functional equation relating the two expressions. We stress, however, that `simplicity' can sometimes be a purely subjective notion -- mathematically it is always exactly the same analytic function!

The first time a striking simplification of a loop amplitude was achieved in ref.~\cite{Goncharov:2010jf}, where also the symbol was introduced for the first time in the physics literature. In ref.~\cite{DelDuca:2009au,DelDuca:2010zg} the so-called two-loop six-point remainder function in $\mathcal{N}=4$ SYM was evaluated, and the result was expressed as a 17-page-long combination of MPLs of uniform weight four. In ref.~\cite{Goncharov:2010jf} the symbol map was used to rewrite the same function
as a single line of classical polylogarithms. By now these techniques have also found there way into QCD computations, and have in particular been used to simplify the analytic expressions for the two-loop amplitudes for a Higgs boson plus three partons~\cite{Gehrmann:2011aa,Brandhuber:2012vm,Duhr:2012fh}, light-quark contributions to top-pair production~\cite{Bonciani:2013ywa} as well as diboson production at two loops~\cite{Gehrmann:2013cxs,Gehrmann:2014bfa}

The first question we may ask is whether there are criteria to ensure the existence of a simpler representation of a function.
In Conjecture~\ref{eq:MaxTrans} we have seen that the coefficient of $\epsilon^k$ in the Laurent expansion of an $L$-loop Feynman diagram is conjectured to have weight at most $2L+k$. The constant term of a two-loop integral can thus have weight at most four. It can be shown (cf. ref.~\cite{Lewin,Goncharov-Volumes,Kellerhals-Volumes}) that MPLs of weight at most three can always be expressed in terms of classical polylogarithms, consistent with the fact that one-loop integrals in four space-time dimensions can always be expressed in terms of dilogarithms and ordinary logarithms. This statement is no longer true for MPLs of weight four. Based on a conjecture in ref.~\cite{Goncharov_delta}, a necessary and sufficient condition was formulated in ref.~\cite{Goncharov:2010jf} to determine whether a given combination of MPLs of weight four can be expressed in terms of classical polylogarithms only. To state this criterion, it is convenient to introduce a linear operator $\delta_4$ acting on tensors of weight four by
\beq
\delta_4(a_1\otimes a_2\otimes a_3\otimes a_4) \equiv (a_1\wedge a_2)\wedge(a_3\wedge a_4)\,,
\eeq
where we defined $a\wedge b\equiv a\otimes b-b\otimes a$. We then have the following
\begin{conjecture}
Let $F$ be a combination of MPLs of weight four. Then $F$ can be expressed in terms of classical polylogarithms only if and only if $\delta_4(\cS(F))=0$.
\end{conjecture}
Similar conjectures can be made for higher weights~\cite{Golden:2014xqa}, and also to decide whether a function can be expressed through product of lower weight functions only~\cite{Ree:1958ab,Duhr:2011zq,Griffing:1995ab}. A detailed discussion of all of these criteria would however go beyond the scope of these lectures.

We have now criteria to decide if a given can be written in a simplified form, but so far we have not answered the question how to actually find this form. In practise, finding this simplified form can be very difficult, even if we know that it exists. The general idea is that if we manage to write down an ansatz for what the simplified form is in terms of MPLs whose coefficients are some unknown rational numbers then we can fix the coefficients by using the Hopf algebra techniques of \sref{duhr_sec:Hopf}. The difficulty, however, often lies in finding this ansatz in the first place, and often this cannot be done without additional input.
One of the issues is related to finding the arguments of the polylogarithms to write down an ansatz. In ref.~\cite{Duhr:2011zq} was presented to find rational arguments that can appear as arguments of polylogarithms with a prescribed alphabet. Once a set of possible arguments is identified, it is sufficient to write down an ansatz of all polylogarithmic functions in these arguments and to fix the coefficients by requiring for example the symbols of the ansatz and the original function to agree. Note that there might not be a unique solution for the coefficients. Indeed, if there are residual relations among the functions appearing in the ansatz, the coefficients can only be fixed up to these relations.

\subsection{Direct integration}
The last application that we are going to discuss is the use of algebraic techniques to explicitly compute integrals. We often led to compute mutlifold integrals over rational functions, and in many cases modern computer algebra systems often fail to do the integrals. The reason is that, if we try to do the integrals one-by-one, the integrand is more complicated after each integration, because every integration in principle increases the weight of the integrand by one unit:
\begin{arabiclist} 
\item The first integrations usually leads to a logarithms.
\item The second integration produces a dilogarithm, i.e., a function of weight two, with a complicated argument.
\item The third integration produces a function of weight three with a complicated argument.
\item etc.
\end{arabiclist}
The idea is then to use at each step functional equations to rewrite the integrand in a form where the next integration can be done using the definition of MPLs, \eref{eq:MPL_def}. MPLs are defined by iterated integration of \emph{linear} factors, and so  at each step in the integration process we need to find a variable in which all the denominators are linear. There are criteria that allow one to determine a priori if there is an order of the integration variables such that this procedure succeeds~\cite{Brown:2008um,Brown:2009ta}. If so, it is possible to perform all the integrations in an algorithmic way~\cite{Brown:2008um} (see also ref.~\cite{Panzer:2013cha,Panzer:2014gra,Panzer:2014caa,Anastasiou:2013srw,Ablinger:2014yaa,Bogner:2014mha}). We note that for this strategy to succeed, it is mandatory to have a convergent integral, so that we can expand in $\eps$ under the integration sign. Extensions of this method to divergent integrals were discussed in ref.~\cite{Panzer:2014gra}.

\begin{example}
Let us illustrate this method on the following integral:
\beq\bsp
\cI(\eps) &\,= \int_0^\infty dx_1dx_2dx_3\,x_1^{\epsilon } \left(1+x_1\right)^{3 \epsilon -2} x_2^{-\epsilon } \left(1+x_2\right)^{-4 \epsilon -2} x_3^{2 \epsilon } \left(1+x_3\right)^{-\epsilon -1} \\
&\,\times\left(1+x_2+x_3+x_1 x_3\right)^{-2 \epsilon -1}\,.
\esp\eeq
Out goal is to compute the first few terms in the $\eps$-expansion of the integral. It is easy to check that the integral
 is finite as $\eps\to0$, and so we can immediately expand in epsilon under the integration sign,
\beq
\cI(\eps) = \cI_0+\cI_1\,\eps+\cI_2\,\eps^2+\cI_3\,\eps^3+\ord(\eps^4)\,.
\eeq
The first coefficient is trivial to compute,
\beq
\cI_0 = \frac{\pi^2}{9}-\frac{2}{3}\,.
\eeq

Next, let us turn to the coefficient $\cI_1$, given by the integral
\beq\bsp
\cI_1&\,=\int_0^\infty \frac{dx_1dx_2dx_3}{\left(1+x_1\right)^2 \left(1+x_2\right)^2 \left(1+x_3\right) \left(1+x_2+x_3+x_1 x_3\right)}\\
&\,\times\Bigg[3 G\left(-1;x_1\right)-4 G\left(-1;x_2\right)-3 G\left(-1;x_3\right)+G\left(0;x_1\right)-G\left(0;x_2\right)\\
&\,+2 G\left(0;x_3\right)-2 G\left(-x_3-1;x_2\right)-2 G\left(\frac{-x_2-x_3-1}{x_3};x_1\right)\Bigg]\,,
\esp\eeq
where we have already written all logarithms in terms of MPLs, e.g.,
\beq\bsp
\log&(1+x_2+x_3+x_1 x_3) \\
&\,= \log(1+x_3) +\log\left(1+\frac{x_2}{1+x_3}\right) + \log\left(1+\frac{x_1\,x_3}{1+x_2+x_3}\right)\\
&\,=G(-1;x_3) + G(-1-x_3;x_2)+G\left(\frac{-x_2-x_3-1}{x_3};x_1\right)\,.
\esp\eeq
It is easy to compute a primitive with respect to $x_1$ for the integrand of $\cI_1$, e.g.,
\beq\label{eq:sampleterm}
\int \frac{dx_1}{1+x_2+x_3+x_1 x_3}G(-1;x_1) = \frac{1}{x_3}G\left(\frac{-x_2-x_3-1}{x_3},-1;x_1\right)\,.
\eeq
In the following we only concentrate on this single term (which is in fact the most complicated one) to illustrate the procedure. All other terms can be dealt with in a similar way. We now need to take the limits $x_1\to0$ and $x_1\to\infty$ of the primitive.
The limit $x_1\to0$ is trivial,
\beq
\lim_{x_1\to0}G\left(\frac{-x_2-x_3-1}{x_3},-1;x_1\right) = 0\,.
\eeq
The limit $x_1\to\infty$ is obtained by letting $x_1=1/\bar{x}_1$ and deriving the inversion relation for this MPL, which can be done using the techniques of \sref{duhr_sec:Hopf}. For a more algorithmic approach we refer to ref.~\cite{Anastasiou:2013srw}.
We find
\beq\bsp\label{eq:liminf}
G&\left(-1,\frac{-x_2-x_3-1}{x_3};x_1\right) = G(0,0;x_1)-G\left(-1;x_2\right) G\left(-1;x_3\right)\\
&\,+G\left(-1;x_2\right) G\left(0;x_3\right)-G\left(-1,-x_3-1;x_2\right)+G\left(0,-1;x_3\right)-G\left(0,0;x_3\right)\\
&\,-\zeta_2+\ord(1/x_1)\,.
\esp\eeq
Note that the function has a logarithmic singularity for $x_1\to\infty$, which will cancel against similar contributions from other terms. The same steps can easily be repeated for all the terms appearing in the primitive with respect to $x_1$. 

After having taken the limits, we can immediately compute the primitive with respect to $x_2$, e.g.,
\beq
\int\frac{dx_2}{1+x_2+x_3}\,G(-1;x_2) = G(-1-x_3,-1;x_2)\,.
\eeq
The limit $x_2\to0$ is again trivial, while the limit $x_2\to\infty$ can again be computed by letting $x_2=1/\bar{x}_2$ and deriving the inversion relation and letting $\bar{x}_2\to 0$. We find
\beq
G(-1-x_3,-1;x_2) = G(0,0;x_2)-G\left(0,-1;x_3\right)+\ord(1/x_2)\,.
\eeq

We are finally only left with the integral over $x_3$. The primitive involves integrals like
\beq
\int\frac{dx_3}{1+x_3}\,G(-1,0;x_3) = G(-1,-1,0;x_3)\,.
\eeq
Proceeding just like before to take the limits, we finally get
\beq
\cI_1 = -5 \zeta_3+\frac{2 \pi ^2}{9}+\frac{5}{3}\,.
\eeq
The higher terms in the $\eps$ expansion can be obtained in exactly the same way. For this particular integral we find for example
\beq\bsp
\cI_2 &\,= \frac{149 \pi ^4}{216}-10 \zeta_3-\frac{16 \pi ^2}{9}-\frac{157}{6}\,,\\
\cI_3&\,= -\frac{910 }{3}\zeta_5+\frac{149 \pi ^4}{108}+\frac{607 }{6}\zeta_3-\frac{277 \pi ^2}{18} \zeta_3+\frac{29 \pi ^2}{3}+\frac{1175}{12}\,.
\esp\eeq
\end{example}


\section{Conclusion}
In these lectures we described mathematical and algebraic structures governing multiple polylogarithms, a class of special functions through which large classes of multi-loop Feynman integrals can be expressed. In particular, we discussed functional equations for multiple polylogarithms and how these relations are governed by the Hopf algebra structure underlying these functions. 

Although progress in understanding the mathematics underlying multi-loop integrals has been fast over the last couple of years and many results have been obtained that were thought impossible only a few years ago, there is still a lot to do. It is known that starting from two loops not every Feynman integrals can be expressed through multiple polylogarithms alone, but generalisations of polylogarithms to elliptic curves appear~\cite{Caffo:1998du,Laporta:2004rb,Bloch:2013tra,Adams:2014vja,Bloch:2014qca}. Currently, only very little is known about these functions, both on the physics and on the mathematics side. Understanding the structure of these functions and how the structures presented in these lectures generalise to higher genus is a fascinating topic, that will most likely lead to new results both in physics and in number theory.


\section{Appendix}\label{duhr_app:algebra}
\subsection{Rings and fields}
A \emph{ring} is a set $R$ equipped with an addition $+$ and a multiplication $\cdot$ such that
\begin{romanlist}
\item $R$ is an additive commutative group.
\item The multiplication is associative and has a unit element.
\item The distributivity law holds:
\beq\bsp
a\cdot (b+c) &\,= a\cdot b+ a\cdot c\,,\\
(a+b)\cdot b &\,= a\cdot b+ a\cdot c\,.
\esp\eeq
\end{romanlist}
Note that the multiplication may be commutative, but this is not mandatory. Moreover, we do not require $R$ to be a multiplicative group, i.e., not every element has a multiplicative inverse. If every element has a multiplicative inverse and the multiplication is commutative, then we call $R$ a \emph{field}. A ring homomorphism is a map $\phi$ between two rings such that the ring structure is preserved, i.e., $\forall a,b\in R$, 
\beq
\phi(a+b) = \phi(a)+\phi(b) {\rm~~and~~} \phi(a\cdot b)=\phi(a)\cdot \phi(b)\,.
\eeq

\subsection{Tensor products and algebras}
Consider two vector spaces $V$ and $W$. The tensor product of $V$ and $W$ is defined by the following universal property: there is a unique vector space (unique up to isomorphism), denoted by $V\otimes W$, together with a bilinear map $\tau:V\times W \to V\otimes W$ such that for every vector space $E$ and every bilinear map $b: V\times W\to E$ there is a unique \emph{linear} map $\beta:V\otimes W\to E$ such that $b = \beta\circ\tau$.

An \emph{algebra} is a vector space $\mathcal{A}$ equipped with a multiplication that turns it into a ring. In other words, there is a map $m:\mathcal{A}\times \mathcal{A}\to \mathcal{A}$ such that $m(a,b)=a\cdot b$. The distributivity law implies that $m$ is bilinear. Then, according to the defining property of the tensor product, there is a linear map $\mu:\mathcal{A}\otimes\mathcal{A}\to\mathcal{A}$ such that $\mu(a\otimes b) = m(a,b)=ab$. In the following, we therefore define an algebra as a vector space $\mathcal{A}$ with a linear map $\mu:\mathcal{A}\otimes\mathcal{A}\to\mathcal{A}$ and a unit element, and the multiplication is associative, 
\beq
\mu(\textrm{id}\otimes\mu) = \mu(\mu\otimes\textrm{id})\,.
\eeq
An algebra homomorphism is a map $\phi$ that preserves the algebra structure, i.e., it is linear and $\phi(a\cdot b)=\phi(a)\cdot\phi(b)$.

If $\mathcal{A}$ and $\mathcal{B}$ are algebras, then their tensor product $\mathcal{A}\otimes\mathcal{B}$ is also an algebra, and the multiplication is given by
\beq
(a_1\otimes b_1)\cdot (a_2\otimes b_2) = (a_1\cdot a_2)\otimes(b_1\cdot b_2)\,.
\eeq

An algebra is called \emph{graded} if it is a direct sum as a vector space
\beq
\mathcal{A} =\bigoplus_{n=0}^\infty\mathcal{A}_n\,,
\eeq and the multiplication preserves the weight
\beq
\mathcal{A}_m\cdot\mathcal{A}_n\subset\mathcal{A}_{m+n}\,.
\eeq

\subsection{Coalgebras and Hopf algebras}
Consider two (complex, real or rational) vector spaces $V$ and $W$ and a linear map $\phi:V\to W$. Our goal is to understand what this map is in terms of the dual spaces. The dual space $W^\ast$ of $W$ is the vector space of all linear functionals $\varphi:W\to\mathbb{K}$ (with $\mathbb{K} = \mathbb{C}$, $\mathbb{R}$ or $\mathbb{Q}$) and which associates to $w\in W$ an element $\varphi(w)\equiv\langle\varphi|w\rangle$. Every linear form is determined by taking the scalar product with a certain constant vector. The hermitian conjugate $\phi^\dagger:W^\ast \to V^\ast$ is defined by 
\beq
\langle \varphi | \phi(v)\rangle \equiv \langle\phi^\dagger(\varphi)|v\rangle\,, \quad \textrm{ for }  v\in V \textrm{ and }\varphi \in W^\ast\,.
\eeq
In particular, if $V=W\otimes W$, we have
\beq
\langle \varphi | \phi(a\otimes b)\rangle \equiv \langle \phi^\dagger(\varphi)|a\otimes b\rangle\,, \quad \textrm{ for }  a,b\in W \textrm{ and }\varphi \in W^\ast\,,
\eeq
where the scalar product in $W\otimes W$ is defined via
\beq
\langle a\otimes b|c\otimes d\rangle \equiv \langle a|c\rangle\,\langle b|d\rangle\,,\quad \textrm{ for }a,b,c,d\in W\,.
\eeq

In the special case where $W$ is an algebra, we have a natural map $\mu:W\otimes W\to W$, and so we may ask what the `hermitian conjugate' $\Delta\equiv \mu^\dagger: W\to W\otimes W$ of the multiplication $\mu$ is. Obviously, $\Delta$ is linear. Writing $\Delta(\varphi) = \sum_i\varphi_i^{(1)}\otimes\varphi_i^{(2)}$, we see that
\beq\bsp
\langle \varphi|(a\cdot b)\cdot c\rangle &\,= \langle \varphi|\mu(\mu(a\otimes b)\otimes c)\rangle\\
&\,=\langle \Delta(\varphi)|\mu(a\otimes b)\otimes c\rangle\\
&\,=\sum_i\langle\varphi_i^{(1)}\otimes\varphi_i^{(2)}|\mu(a\otimes b)\otimes c\rangle\\
&\,=\sum_i\langle\varphi_i^{(1)}|\mu(a\otimes b)\rangle\,\langle\varphi_i^{(2)}|c\rangle\\
&\,=\sum_i\langle\Delta(\varphi_i^{(1)})|a\otimes b\rangle\,\langle\varphi_i^{(2)}|c\rangle\\
&\,=\langle(\Delta\otimes \textrm{id})\Delta(\varphi)|a\otimes b\otimes c\rangle\,.
\esp\eeq
Similarly, we get $\langle \varphi|a\cdot (b\cdot c)\rangle = \langle(\textrm{id}\otimes\Delta)\Delta(\varphi)|a\otimes b\otimes c\rangle$, and the associativity of $\mu$ implies that these two expressions must be equal, and so we must have
\beq
(\Delta\otimes \textrm{id})\Delta = (\textrm{id}\otimes\Delta)\Delta\,,
\eeq
i.e., we see that $\Delta$ is coassociative. In other words, if $W$ is an algebra then $W^\ast$ is a coalgebra\footnote{There is also a similar dual notion of the unit element of $W$.}.

A coalgebra $W$ is called graded if it is a direct sum of vector spaces,
\beq
W=\bigoplus_{n=0}^\infty W_n\,,
\eeq
and the coproduct preserves the weight
\beq
\Delta(W_n)\subset \bigoplus_{k=0}^nW_k\otimes W_{n-k}\,.
\eeq

A \emph{bialgebra} is an algebra that is at the same time a coalgebra, and the product and the coproduct are compatible in the sense that
\beq
\Delta(a\cdot b) = \Delta(a)\cdot \Delta(b)\,,
\eeq
i.e., the coproduct is an algebra homomorphism.

A \emph{Hopf algebra} $H$ is a bialgebra equipped with an antipode, a linear map $S:H\to H$ satisfying certain properties. It turns out that in our case the antipode does not contain new information because one can show that if a bialgebra $H$ is graded and $H_0=\mathbb{Q}$, then there is a unique antipode that turns $H$ into a Hopf algebra. In other words, the antipode does not carry any information that was not already present at the level of the bialgebra, and we therefore never consider it explicitly in these lectures.

\section*{Acknowledgements}
I would each like the thank the TASI organisers (Lance Dixon and Frank Petriello) for inviting me to lecture at the TASI school 2014 and for creating such a
stimulating environment for students and lecturers alike. I am grateful to Nicolas Deutschmann for comments on the manuscript. This work is supported by the ``Fonds National de la Recherche Scientifique'' (FNRS), Belgium.

\bibliographystyle{ws-rv-van}
\bibliography{Duhr_TASI}
\end{document}